\documentclass[journal,draftcls,onecolumn,12pt,twoside]{IEEEtranTCOM}
\normalsize
\ifCLASSINFOpdf
\else
\fi
\usepackage{lineno}
\usepackage{cite}
\usepackage{hyperref}
\usepackage{amsmath,amssymb,amsfonts}
\usepackage{amsmath}
\usepackage{amsthm}

\usepackage{graphicx}
\usepackage{textcomp}
\usepackage{xcolor}
\usepackage{graphicx}
\usepackage{float}
\usepackage{subfigure}
\usepackage{amsmath}
\usepackage{amsfonts,amssymb}
\usepackage{mathrsfs}
\usepackage{mathtools}
\usepackage{algorithm}
\usepackage{algorithmicx}
\usepackage{algpseudocode}
\usepackage{bm}
\usepackage{multirow}
\usepackage{array}
\usepackage{amssymb}
\usepackage{amsmath}
\usepackage{cite}
\usepackage{url}
\usepackage{xcolor}
\usepackage{cite,graphicx,amsmath,amssymb}
\usepackage{subfigure}
\usepackage{fancyhdr}
\usepackage{mdwmath}
\usepackage{mdwtab}
\usepackage{caption}
\usepackage{amsthm}
\usepackage{setspace}
\usepackage{bm}
\usepackage{algorithm}
\usepackage{algpseudocode}
\usepackage{mathtools}
\usepackage{dsfont}
\usepackage{bbm}

\newtheorem{theorem}{Theorem}

\newtheorem{lemma}{Lemma}

\newtheorem{corollary}{Corollary}

\makeatletter
\newcommand{\biggg}{\bBigg@{3}}
\newcommand{\Biggg}{\bBigg@{3.5}}
\makeatother
\usepackage{bm}

\begin{document}
\title{MMSE Bound for MIMO Channel\\
}

\author{Chongjun~Ouyang and Hongwen~Yang
\thanks{C. Ouyang and H. Yang are with the School of Information and Communication Engineering, Beijing University of Posts and Telecommunications, Beijing, 100876, China (e-mail: \{DragonAim,~yanghong\}@bupt.edu.cn).}}
\maketitle
\vspace{-70pt}

\begin{abstract}
Detailed derivations of two bounds of the minimum mean-square error (MMSE) of complex-valued multiple-input multiple-output (MIMO) systems are proposed for performance evaluation. Particularly, the lower bound is derived based on a genie-aided MMSE estimator, whereas the upper bound is derived based on a maximum-likelihood (ML) estimator. Using the famous relationship between the mutual information (MI) and MMSE, two bounds for the MI are also derived, based on which we discuss the asymptotic behaviours of the average MI in the high-signal-to-noise ratio (SNR) regime. Theoretical analyses suggest that the average MI will converge its maximum as the SNR increases and the diversity order is the same as receive antenna number.
\end{abstract}

\begin{IEEEkeywords}
Minimum mean-square error (MMSE), mutual information (MI), performance bounds. 	
\end{IEEEkeywords}

\section{Introduction}
In 2005, Guo \emph{et al.} explored the basic relationship between the minimum mean-square error (MMSE) and mutual information (MI) of multiple-input multiple-output (MIMO) systems, which states that the derivative of the MI with respect to the signal-to-noise ratio (SNR) equals the corresponding MMSE \cite{Guo2005}. Furthermore, the complex gradient of the MI with respect to the precoding matrices were derived for MIMO channels and a numerical precoding method was proposed for MIMO channel with finite inputs \cite{Palomar2006}. Using these results, the authors discussed the power allocation policy among parallel sub-channels with arbitrary inputs and discuused several properties of the MMSE of a single-input single-output (SISO) channel \cite{Lozano2006}. Later, the authors in \cite{Rodrigues2010} tried to provide an optimal linear precoding and power allocation policies that maximize the mutual information based on necessary but not sufficient conditions. Yet, the global optimum of these algorithms cannot be guaranteed. As a compromise, the optimal precoding method for MIMO channels with finite-input signals were proposed and its global optimum was strictly proved \cite{Xiao2011}. The works in \cite{Rodrigues2014} further extended the work in \cite{Rodrigues2010} to fading channels and use two bounds to evaluate the asymptotic mutual information of MIMO fading channels. 

Unfortunately, the bounds derived in \cite{Rodrigues2014} was wrong and these bounds were directly presented without detailed discussions. For performance evaluation in MIMO channels with finite inputs, it is necessary to derive the correct bounds for MIMO systems and provide the detailed steps. To solve this problem, this paper derived tow correct bounds for MIMO channel with arbitrary inputs. Besides, we use these bounds to analyze the diversity order of the average mutual information over MIMO fading channels. Specifically, we will first discuss the SISO case, then the single-input multiple-output (SIMO) case, and finally the MIMO case.

\section{Single-Input Single-Output (SISO)}
Consider a real-valued SISO channel given as
\begin{align}\label{SISO_Model}
y=\sqrt{\mathsf{snr}}x+n,
\end{align}
where $x$ denotes the transmitted symbol satisfying $\mathbbmss{E}\left\{\left|x\right|^2\right\}=1$, $\mathsf{snr}>0$ denotes the signal-to-noise ratio (SNR), and $n\sim{\mathcal{N}}\left(0,1\right)$ denotes the additive white Gaussian noise. The input-output mutual information (MI) of the SISO channel can be written as
\begin{align}
{\mathsf{I}}_{\text{siso}}\left(\mathsf{snr}\right)={\mathbb{E}}\left\{\left.\log\left(\frac{p\left(\left.x,y\right|\mathsf{snr}\right)}
{p\left(\left.x\right|\mathsf{snr}\right)p\left(\left.y\right|\mathsf{snr}\right)}\right)\right|\mathsf{snr}\right\},
\end{align}
where $p\left(\left.x,y\right|\mathsf{snr}\right)$, $p\left(\left.x\right|\mathsf{snr}\right)$, and $p\left(\left.y\right|\mathsf{snr}\right)$ denote the joint probability density function (PDF) of $\left(x,y\right)$ given $\mathsf{snr}$, the PDF of $x$ given $\mathsf{snr}$, and the PDF of $y$ given $\mathsf{snr}$, respectively. The error associated with the estimation of the noiseless input, $x$, given the noisy output, $y$, of the channel can be measured in mean-square sense
\begin{align}
{\mathsf{mse}}=\mathbbmss{E}\left\{\left.\left(x-f\left(y\right)\right)^2\right|y\right\},
\end{align}
where $f\left(y\right)$ denotes the estimation. Since the output, $y$, is given, both $y$ and $f\left(y\right)$ can be treated as constants. Therefore, we have
\begin{align}
{\mathsf{mse}}&=\mathbbmss{E}\left\{\left.\left(x-f\left(y\right)\right)^2\right|y\right\}\\
&=\mathbbmss{E}\left\{\left.x^2-2xf\left(y\right)+f^2\left(y\right)\right|y\right\},\\
&=\mathbbmss{E}\left\{\left.x^2\right|y\right\}-2f\left(y\right)\mathbbmss{E}\left\{\left.x\right|y\right\}+f^2\left(y\right).
\end{align}
By treating the mean-square error (MSE) as a function of $f\left(y\right)$ and ignoring the terms independent of it, we find the optimal estimator that can achieve the minimum value of ${\mathsf{mse}}$, referred to as the minimum MSE (MMSE), is the conditional mean estimator
\begin{align}
f^{\star}\left(y\right)=\arg\min_{f\left(y\right)}=\mathbbmss{E}\left\{\left.x\right|y\right\}.
\end{align}
The associated with the MMSE estimator is given by
\begin{align}
{\mathsf{mmse}}_{\text{siso}}\left(\mathsf{snr}\right)=\mathbbmss{E}\left\{\left.\left(x-\mathbbmss{E}\left\{\left.x\right|y\right\}\right)^2\right|\mathsf{snr}\right\},
\end{align}
where the expectation is taken over $\left(x,y\right)$. Based on the famous MI-MMSE relationship, we have
\begin{align}
\frac{{\rm{d}}}{{\rm{d}}{\mathsf{snr}}}{\mathsf{I}}_{\text{siso}}\left(\mathsf{snr}\right)=\frac{1}{2}
{\mathsf{mmse}}_{\text{siso}}\left(\mathsf{snr}\right).
\end{align}
In the sequel, we consider a special case where $x$ in \eqref{SISO_Model} is taken from a BPSK (binary phase shift keying) constellation alphabet $\left\{-1,+1\right\}$ with equal probability. Under this circumstance, the corresponding MMSE estimator can be written as
\begin{align}
\mathbbmss{E}\left\{\left.x\right|y\right\}=\int_{-\infty}^{+\infty} xp\left(x|y,\mathsf{snr}\right){\rm{d}}x,
\end{align}
where $p\left(x|y,\mathsf{snr}\right)$ denotes the conditioned PDF of $x$ given $y$ and $\mathsf{snr}$. According to Bayes' theorem, we have
\begin{align}
p\left(x|y,\mathsf{snr}\right)=\frac{p\left(\left.x,y\right|\mathsf{snr}\right)}{p\left(\left.y\right|\mathsf{snr}\right)},
\end{align}
where the joint PDF of $\left(x,y\right)$ for a given $\mathsf{snr}$ can be written as
\begin{equation}
\begin{split}
p\left(\left.x,y\right|\mathsf{snr}\right)&=\frac{1}{2}\delta\left(x-1\right)\frac{1}{\sqrt{2\pi}}\exp\left(-\frac{\left(y-\sqrt{\mathsf{snr}}\right)^2}{2}\right)\\
&+\frac{1}{2}\delta\left(x+1\right)\frac{1}{\sqrt{2\pi}}\exp\left(-\frac{\left(y+\sqrt{\mathsf{snr}}\right)^2}{2}\right)
\end{split}
\end{equation}
with $\delta\left(\cdot\right)$ denoting the Dirac delta function, and where the PDF of $y$ for a given $\mathsf{snr}$ is given as
\begin{align}
p\left(\left.y\right|\mathsf{snr}\right)&=\Pr\left(x=1\right)p\left(\left.y\right|x=1,\mathsf{snr}\right)+\Pr\left(x=-1\right)p\left(\left.y\right|x=-1,\mathsf{snr}\right)\\
&=\frac{1}{2}\frac{1}{\sqrt{2\pi}}\exp\left(-\frac{\left(y-\sqrt{\mathsf{snr}}\right)^2}{2}\right)+
\frac{1}{2}\frac{1}{\sqrt{2\pi}}\exp\left(-\frac{\left(y+\sqrt{\mathsf{snr}}\right)^2}{2}\right)
\end{align}
with $\Pr\left(x=a\right)$ denotes the probability of event $x=a$. Taken together, we can obtain
\begin{align}
\mathbbmss{E}\left\{\left.x\right|y\right\}&=\int_{-\infty}^{+\infty} xp\left(x|y,\mathsf{snr}\right){\rm{d}}x
=\frac{\frac{1}{2}\frac{1}{\sqrt{2\pi}}\exp\left(-\frac{\left(y-\sqrt{\mathsf{snr}}\right)^2}{2}\right)-
\frac{1}{2}\frac{1}{\sqrt{2\pi}}\exp\left(-\frac{\left(y+\sqrt{\mathsf{snr}}\right)^2}{2}\right)}
{\frac{1}{2}\frac{1}{\sqrt{2\pi}}\exp\left(-\frac{\left(y-\sqrt{\mathsf{snr}}\right)^2}{2}\right)+
\frac{1}{2}\frac{1}{\sqrt{2\pi}}\exp\left(-\frac{\left(y+\sqrt{\mathsf{snr}}\right)^2}{2}\right)}\\
&=\frac{\exp\left(-\frac{\left(y-\sqrt{\mathsf{snr}}\right)^2}{2}\right)-
\exp\left(-\frac{\left(y+\sqrt{\mathsf{snr}}\right)^2}{2}\right)}
{\exp\left(-\frac{\left(y-\sqrt{\mathsf{snr}}\right)^2}{2}\right)+
\exp\left(-\frac{\left(y+\sqrt{\mathsf{snr}}\right)^2}{2}\right)}
=\frac{\exp\left(\sqrt{\mathsf{snr}}y\right)-
\exp\left(-\sqrt{\mathsf{snr}}y\right)}
{\exp\left(\sqrt{\mathsf{snr}}y\right)+
\exp\left(-\sqrt{\mathsf{snr}}y\right)}\\
&=\tanh\left(\sqrt{\mathsf{snr}}y\right),
\end{align}
where $\tanh\left(x\right)=\frac{\exp\left(x\right)-
\exp\left(-x\right)}
{\exp\left(x\right)+
\exp\left(-x\right)}$ represents the hyperbolic tangent function. In the sequel, we derive the MSE achieved by this estimator, namely the MMSE, which is given by
\begin{align}
{\mathsf{mmse}}_{\text{siso}}\left(\mathsf{snr}\right)=\mathbbmss{E}\left\{\left.\left(x-\tanh\left(\sqrt{\mathsf{snr}}y\right)\right)^2\right|\mathsf{snr}\right\},
\end{align}
where the expectation is taken over $\left(x,y\right)$. Therefore, the MMSE can be calculated as
\begin{align}
{\mathsf{mmse}}_{\text{siso}}\left(\mathsf{snr}\right)&=
\int_{-\infty}^{+\infty}\int_{-\infty}^{+\infty}\left(x-\tanh\left(\sqrt{\mathsf{snr}}y\right)\right)^2p\left(\left.x,y\right|\mathsf{snr}\right){\rm{d}}x{\rm{d}}y\\
&=\frac{1}{2}\int_{-\infty}^{+\infty}\left(1-\tanh\left(\sqrt{\mathsf{snr}}y\right)\right)^2\frac{1}{\sqrt{2\pi}}\exp\left(-\frac{\left(y-\sqrt{\mathsf{snr}}\right)^2}{2}\right)\nonumber\\
&+\frac{1}{2}\int_{-\infty}^{+\infty}\left(1+\tanh\left(\sqrt{\mathsf{snr}}y\right)\right)^2\frac{1}{\sqrt{2\pi}}\exp\left(-\frac{\left(y+\sqrt{\mathsf{snr}}\right)^2}{2}\right)\\
&=1-\int_{-\infty}^{+\infty}\frac{\tanh\left(\sqrt{\mathsf{snr}}y\right)}{\sqrt{2\pi}}\left(
\exp\left(-\frac{\left(y-\sqrt{\mathsf{snr}}\right)^2}{2}\right)-\exp\left(-\frac{\left(y+\sqrt{\mathsf{snr}}\right)^2}{2}\right)
\right)\nonumber\\
&+\int_{-\infty}^{+\infty}\frac{\tanh^2\left(\sqrt{\mathsf{snr}}y\right)}{2\sqrt{2\pi}}\left(
\exp\left(-\frac{\left(y-\sqrt{\mathsf{snr}}\right)^2}{2}\right)+\exp\left(-\frac{\left(y+\sqrt{\mathsf{snr}}\right)^2}{2}\right)
\right).
\end{align}
Substituting $\tanh\left(\sqrt{\mathsf{snr}}y\right)=\frac{\exp\left(-\frac{\left(y-\sqrt{\mathsf{snr}}\right)^2}{2}\right)-
\exp\left(-\frac{\left(y+\sqrt{\mathsf{snr}}\right)^2}{2}\right)}
{\exp\left(-\frac{\left(y-\sqrt{\mathsf{snr}}\right)^2}{2}\right)+
\exp\left(-\frac{\left(y+\sqrt{\mathsf{snr}}\right)^2}{2}\right)}$ back to the above expression gives
\begin{align}
{\mathsf{mmse}}_{\text{siso}}\left(\mathsf{snr}\right)&=1-\int_{-\infty}^{+\infty}{\frac{\left(\exp\left(-\frac{\left(y-\sqrt{\mathsf{snr}}\right)^2}{2}\right)-
\exp\left(-\frac{\left(y+\sqrt{\mathsf{snr}}\right)^2}{2}\right)\right)^2}
{\exp\left(-\frac{\left(y-\sqrt{\mathsf{snr}}\right)^2}{2}\right)+
\exp\left(-\frac{\left(y+\sqrt{\mathsf{snr}}\right)^2}{2}\right)}}\frac{1}{2\sqrt{2\pi}}{\rm{d}}y\\
&=1-\int_{-\infty}^{+\infty}{\tanh\left(\sqrt{\mathsf{snr}}y\right)\frac{\exp\left(-\frac{\left(y-\sqrt{\mathsf{snr}}\right)^2}{2}\right)-
\exp\left(-\frac{\left(y+\sqrt{\mathsf{snr}}\right)^2}{2}\right)}{2\sqrt{2\pi}}{\rm{d}}y}.\label{SISO_Real_MMSE_Basic}
\end{align}
Note that the hyperbolic tangent function is an odd function, and thus $\tanh\left(x\right)=-\tanh\left(-x\right)$. On this basis,
\begin{equation}\label{SISO_Real_Important_Trans}
\begin{split}
&\int_{-\infty}^{+\infty}{\tanh\left(\sqrt{\mathsf{snr}}y\right)
\exp\left(-\frac{\left(y+\sqrt{\mathsf{snr}}\right)^2}{2}\right)}{\rm{d}}y\\&=
\int_{-\infty}^{+\infty}{\tanh\left(-\sqrt{\mathsf{snr}}y\right)
\exp\left(-\frac{\left(-y+\sqrt{\mathsf{snr}}\right)^2}{2}\right)}{\rm{d}}y\\&=
-\int_{-\infty}^{+\infty}{\tanh\left(\sqrt{\mathsf{snr}}y\right)
\exp\left(-\frac{\left(y-\sqrt{\mathsf{snr}}\right)^2}{2}\right)}{\rm{d}}y,
\end{split}
\end{equation}
where the first equality is due to the variable change $y\rightarrow-y$ and the second equality is due to the facts of $\left(-y+\sqrt{\mathsf{snr}}\right)^2=\left(y-\sqrt{\mathsf{snr}}\right)^2$ and $-\tanh\left(\sqrt{\mathsf{snr}}y\right)=\tanh\left(-\sqrt{\mathsf{snr}}y\right)$. We then inserting \eqref{SISO_Real_Important_Trans} into \eqref{SISO_Real_MMSE_Basic} and obtain
\begin{align}
{\mathsf{mmse}}_{\text{siso}}\left(\mathsf{snr}\right)
=1-\int_{-\infty}^{+\infty}{\tanh\left(\sqrt{\mathsf{snr}}y\right)\frac{\exp\left(-\frac{\left(y-\sqrt{\mathsf{snr}}\right)^2}{2}\right)}{\sqrt{2\pi}}{\rm{d}}y}.
\end{align}

We now move to a more general case where the transmitted symbol, $x$, in \eqref{SISO_Model} is taken from a complex constellation alphabet and the additive noise follows complex Gaussian distribution, namely $n\sim{\mathcal{CN}}\left(0,1\right)$. In this case, the MSE achieved by estimator, $f\left(y\right)$, is given by
\begin{align}
{\mathsf{mse}}&=\mathbbmss{E}\left\{\left.\left|x-f\left(y\right)\right|^2\right|y\right\}\\
&=\mathbbmss{E}\left\{\left.xx^\dag\right|y\right\}-f^{\dag}\left(y\right)\mathbbmss{E}\left\{\left.x\right|y\right\}-
f\left(y\right)\mathbbmss{E}\left\{\left.x^{\dag}\right|y\right\}
+f\left(y\right)f^{\dag}\left(y\right),
\end{align}
where $\left(\cdot\right)^\dag$ denotes the conjugate transpose operator. By calculating the complex gradient of ${\mathsf{mse}}$ with respect to $f\left(y\right)$ and then setting it to zero, we have
\begin{align}
\mathbbmss{E}\left\{\left.x\right|y\right\}-f\left(y\right)=0,
\end{align}
which suggests that the MMSE estimator is given by
\begin{align}
f^{\star}\left(y\right)=\arg\min_{f\left(y\right)}=\mathbbmss{E}\left\{\left.x\right|y\right\}.
\end{align}
Therefore, the corresponding MMSE can be expressed as
\begin{align}
{\mathsf{mmse}}_{\text{siso}}\left(\mathsf{snr}\right)=\int\int\left|x-\mathbbmss{E}\left\{\left.x\right|y\right\}\right|^2p\left(\left.x,y\right|\mathsf{snr}\right){\rm{d}}x{\rm{d}}y.
\end{align}
In this complex-value SISO channel, the MI-MMSE relationship satisfies
\begin{align}
\frac{{\rm{d}}}{{\rm{d}}{\mathsf{snr}}}{\mathsf{I}}_{\text{siso}}\left(\mathsf{snr}\right)=
{\mathsf{mmse}}_{\text{siso}}\left(\mathsf{snr}\right).
\end{align}
Assume that the transmitted symbol is taken from a constellation alphabet consisting of $M$ constellation points $\left\{x_1,\cdots,x_M\right\}$ and the probability of $x$ taking $x_i$ is given by $\Pr\left(x=x_i\right)=p_i$. Note that $\sum_{i=1}^{M}p_i=1$. The joint PDF of $\left(x,y\right)$ for a given $\mathsf{snr}$ can be written as
\begin{equation}
\begin{split}
p\left(\left.x,y\right|\mathsf{snr}\right)=\sum_{i=1}^{M}p_i\delta\left(x-x_i\right)
\frac{1}{\pi}\exp\left(-{\left|y-\sqrt{\mathsf{snr}}x_i\right|^2}\right),
\end{split}
\end{equation}
and the PDF of $y$ for a given $\mathsf{snr}$ is given as
\begin{align}
p\left(\left.y\right|\mathsf{snr}\right)=\sum_{i=1}^{M}p_i
\frac{1}{\pi}\exp\left(-{\left|y-\sqrt{\mathsf{snr}}x_i\right|^2}\right).
\end{align}
As a result, the MMSE estimator can be expressed as
\begin{align}
\mathbbmss{E}\left\{\left.x\right|y\right\}=\int xp\left(x|y,\mathsf{snr}\right){\rm{d}}x
=\frac{\sum_{i=1}^{M}p_ix_i
\frac{1}{\pi}\exp\left(-{\left|y-\sqrt{\mathsf{snr}}x_i\right|^2}\right)}
{\sum_{i=1}^{M}p_i
\frac{1}{\pi}\exp\left(-{\left|y-\sqrt{\mathsf{snr}}x_i\right|^2}\right)}\triangleq{\mathcal{S}}_{\text{siso}}\left(y\right).
\end{align}
Accordingly, the MMSE can be calculated as
\begin{align}
{\mathsf{mmse}}_{\text{siso}}\left(\mathsf{snr}\right)=\mathbbmss{E}\left\{\left.\left|x-\frac{\sum_{i=1}^{M}p_ix_i
\frac{1}{\pi}\exp\left(-{\left|y-\sqrt{\mathsf{snr}}x_i\right|^2}\right)}
{\sum_{i=1}^{M}p_i
\frac{1}{\pi}\exp\left(-{\left|y-\sqrt{\mathsf{snr}}x_i\right|^2}\right)}\right|^2\right|\mathsf{snr}\right\},
\end{align}
where the expectation is taken over $\left(x,y\right)$. Therefore, the MMSE can be calculated as
\begin{align}
&{\mathsf{mmse}}_{\text{siso}}\left(\mathsf{snr}\right)=
\int\int\left|x-{\mathcal{S}}_{\text{siso}}\left(y\right)\right|^2p\left(\left.x,y\right|\mathsf{snr}\right){\rm{d}}x{\rm{d}}y\\
&=
\int\sum_{i=1}^{M}p_i\left|x_i-{\mathcal{S}}_{\text{siso}}\left(y\right)\right|^2
\frac{\exp\left(-{\left|y-\sqrt{\mathsf{snr}}x_i\right|^2}\right)}{\pi}{\rm{d}}y\\
&=\sum_{i=1}^{M}p_i\left|x_i\right|^2\int\frac{1}{\pi}{\exp\left(-{\left|y-\sqrt{\mathsf{snr}}x_i\right|^2}\right)}{\rm{d}}y
+\int\left|{\mathcal{S}}_{\text{siso}}\left(y\right)\right|^2\frac{\sum_{i=1}^{M}p_i\exp\left(-{\left|y-\sqrt{\mathsf{snr}}x_i\right|^2}\right)}{\pi}{\rm{d}}y
\nonumber\\
&-\int{\mathcal{S}}_{\text{siso}}\left(y\right)\frac{\sum\limits_{i=1}^{M}x_i^{\dag}p_i\exp\left(-{\left|y-\sqrt{\mathsf{snr}}x_i\right|^2}\right)}{\pi}{\rm{d}}y
-\int{\mathcal{S}}_{\text{siso}}^{\dag}\left(y\right)\frac{\sum\limits_{i=1}^{M}x_ip_i\exp\left(-{\left|y-\sqrt{\mathsf{snr}}x_i\right|^2}\right)}{\pi}{\rm{d}}y
.
\end{align}
It is easy to verify that
\begin{align}
{\mathcal{S}}_{\text{siso}}\left(y\right)\sum\limits_{i=1}^{M}x_i^{\dag}p_i\exp\left(-{\left|y-\sqrt{\mathsf{snr}}x_i\right|^2}\right)
&=
\frac{\left|\sum_{i=1}^{M}p_ix_i
\exp\left(-{\left|y-\sqrt{\mathsf{snr}}x_i\right|^2}\right)\right|^2}
{\sum_{i=1}^{M}p_i
\exp\left(-{\left|y-\sqrt{\mathsf{snr}}x_i\right|^2}\right)}\\
&={\mathcal{S}}_{\text{siso}}^{\dag}\left(y\right)\sum\limits_{i=1}^{M}x_ip_i\exp\left(-{\left|y-\sqrt{\mathsf{snr}}x_i\right|^2}\right)\\
&=\left|{\mathcal{S}}_{\text{siso}}\left(y\right)\right|^2\sum_{i=1}^{M}p_i\exp\left(-{\left|y-\sqrt{\mathsf{snr}}x_i\right|^2}\right)
\end{align}
and
\begin{align}
\sum_{i=1}^{M}p_i\left|x_i\right|^2=\mathbbmss{E}\left\{\left|x\right|^2\right\}=1
\end{align}
According to the above expressions, we can obtain
\begin{align}\label{MMSE_SISO_Complex}
{\mathsf{mmse}}_{\text{siso}}\left(\mathsf{snr}\right)=1-\frac{1}{\pi}\int\frac{\left|\sum_{i=1}^{M}p_ix_i
\exp\left(-{\left|y-\sqrt{\mathsf{snr}}x_i\right|^2}\right)\right|^2}
{\sum_{i=1}^{M}p_i
\exp\left(-{\left|y-\sqrt{\mathsf{snr}}x_i\right|^2}\right)}{\rm{d}}y.
\end{align}
Then, we specialize the above expression to a special case where the BPSK is utilized. Under this circumstance, we have
\begin{align}
{\mathsf{mmse}}_{\text{siso}}\left(\mathsf{snr}\right)=1-\int\tanh\left(2\sqrt{\mathsf{snr}}\Re\left\{y\right\}\right)
\frac{\exp\left(-{\left|y-\sqrt{\mathsf{snr}}\right|^2}\right)-\exp\left(-{\left|y+\sqrt{\mathsf{snr}}\right|^2}\right)}{2\pi}
{\rm{d}}y.
\end{align}
Leveraging the property of $\tanh\left(-x\right)=-\tan\left(x\right)$, we can obtain
\begin{align}
{\mathsf{mmse}}_{\text{siso}}\left(\mathsf{snr}\right)=1-\int\tanh\left(2\sqrt{\mathsf{snr}}\Re\left\{y\right\}\right)
\frac{\exp\left(-{\left|y-\sqrt{\mathsf{snr}}\right|^2}\right)}{\pi}
{\rm{d}}y.
\end{align}
Since the real part and imagine part of $y$ are mutually independent, we can rewrite the above integral as
\begin{align}
{\mathsf{mmse}}_{\text{siso}}\left(\mathsf{snr}\right)=1-\int_{-\infty}^{+\infty}\int_{-\infty}^{+\infty}
\tanh\left(2\sqrt{\mathsf{snr}}a\right)
\frac{\exp\left(-{\left(a-\sqrt{\mathsf{snr}}\right)^2}-b^2\right)}{\pi}
{\rm{d}}a{\rm{d}}b,
\end{align}
where $a=\Re\left\{y\right\}$ and $b=\Im\left\{y\right\}$. Then, we have
\begin{align}
{\mathsf{mmse}}_{\text{siso}}\left(\mathsf{snr}\right)&=1-\int_{-\infty}^{+\infty}
\tanh\left(2\sqrt{\mathsf{snr}}a\right)
\frac{\exp\left(-{\left(a-\sqrt{\mathsf{snr}}\right)^2}\right)}{\sqrt{\pi}}
{\rm{d}}a\int_{-\infty}^{+\infty}\frac{\exp\left(-b^2\right)}{\sqrt{\pi}}{\rm{d}}b\\
&=1-\int_{-\infty}^{+\infty}
\tanh\left(2\sqrt{\mathsf{snr}}a\right)
\frac{\exp\left(-{\left(a-\sqrt{\mathsf{snr}}\right)^2}\right)}{\sqrt{\pi}}
{\rm{d}}a.
\end{align}

\section{Single-Input Multiple-Output (SIMO)}
We now turn our attention to the SIMO channel characterized as follows
\begin{align}
{\textbf{y}}=\sqrt{\mathsf{snr}}{\textbf{h}}x+{\textbf{n}},
\end{align}
where $x$ denotes the transmitted symbol satisfying $\mathbbmss{E}\left\{\left|x\right|^2\right\}=1$, $\mathsf{snr}>0$ denotes the SNR, ${\textbf{h}}\in{\mathbb{C}}^{N\times1}$ denotes the channel vector, and $\textbf{n}\sim{\mathcal{CN}}\left({\textbf{0}},{\textbf{I}}\right)$ denotes the additive white Gaussian noise. The MSE achieved by estimator, $f\left(\textbf{y}\right)$, is given by
\begin{align}
{\mathsf{mse}}&=\mathbbmss{E}\left\{\left.\left\|{\textbf{h}}\left(x-f\left(\textbf{y}\right)\right)\right\|^2\right|\textbf{y}\right\}
=\mathbbmss{E}\left\{\left.\left\|{\textbf{h}}\right\|^2\left|x-f\left(\textbf{y}\right)\right|^2\right|\textbf{y}\right\}\\
&=\left\|{\textbf{h}}\right\|^2\mathbbmss{E}\left\{\left.\left|x-f\left(\textbf{y}\right)\right|^2\right|\textbf{y}\right\}\\
&=\left\|{\textbf{h}}\right\|^2\left(\mathbbmss{E}\left\{\left.xx^\dag\right|\textbf{y}\right\}-f^{\dag}\left(\textbf{y}\right)\mathbbmss{E}\left\{\left.x\right|\textbf{y}\right\}-
f\left(y\right)\mathbbmss{E}\left\{\left.x^{\dag}\right|\textbf{y}\right\}
+f\left(y\right)f^{\dag}\left(\textbf{y}\right)\right).
\end{align}
Upon calculating the complex gradient of ${\mathsf{mse}}$ with respect to $f\left(\textbf{y}\right)$ and then setting it to zero, we can obtain the MMSE estimator as follows
\begin{align}
f^{\star}\left(\textbf{y}\right)=\arg\min_{f\left(\textbf{y}\right)}=\mathbbmss{E}\left\{\left.x\right|\textbf{y}\right\}.
\end{align}
Consequently, the corresponding MMSE can be expressed as
\begin{align}
{\mathsf{mmse}}_{\text{simo}}\left(\mathsf{snr};{\textbf{h}}\right)=\left\|{\textbf{h}}\right\|^2
\int\int\left|x-\mathbbmss{E}\left\{\left.x\right|\textbf{y}\right\}\right|^2p\left(\left.x,\textbf{y}\right|\mathsf{snr}\right){\rm{d}}x{\rm{d}}\textbf{y}.
\end{align}
In this complex-value SIMO channel, the MI-MMSE relationship satisfies
\begin{align}
\frac{{\rm{d}}}{{\rm{d}}{\mathsf{snr}}}{\mathsf{I}}_{\text{simo}}\left(\mathsf{snr};{\textbf{h}}\right)=
{\mathsf{mmse}}_{\text{simo}}\left(\mathsf{snr};{\textbf{h}}\right),
\end{align}
where
\begin{align}
{\mathsf{I}}_{\text{simo}}\left(\mathsf{snr};{\textbf{h}}\right)={\mathbb{E}}\left\{\left.\log\left(\frac{p\left(\left.x,\textbf{y}\right|\mathsf{snr},{\textbf{h}}\right)}
{p\left(\left.x\right|\mathsf{snr},{\textbf{h}}\right)p\left(\left.\textbf{y}\right|\mathsf{snr},{\textbf{h}}\right)}\right)\right|\mathsf{snr},{\textbf{h}}\right\}
\end{align}
denotes the mutual information of this channel. Assume that the transmitted symbol is taken from a constellation alphabet consisting of $M$ constellation points $\left\{x_1,\cdots,x_M\right\}$ and the probability of $x$ taking $x_i$ is given by $\Pr\left(x=x_i\right)=p_i$. Note that $\sum_{i=1}^{M}p_i=1$. The joint PDF of $\left(x,\textbf{y}\right)$ for a given $\mathsf{snr}$ can be written as
\begin{equation}
\begin{split}
p\left(\left.x,\textbf{y}\right|\mathsf{snr},{\textbf{h}}\right)=\sum_{i=1}^{M}p_i\delta\left(x-x_i\right)
\frac{1}{\pi^N}\exp\left(-{\left\|\textbf{y}-\sqrt{\mathsf{snr}}x_i{\textbf{h}}\right\|^2}\right),
\end{split}
\end{equation}
and the PDF of $\textbf{y}$ for a given $\mathsf{snr}$ can be given as
\begin{align}
p\left(\left.\textbf{y}\right|\mathsf{snr},{\textbf{h}}\right)=\sum_{i=1}^{M}p_i
\frac{1}{\pi^N}\exp\left(-{\left\|\textbf{y}-\sqrt{\mathsf{snr}}x_i{\textbf{h}}\right\|^2}\right).
\end{align}
As a result, the MMSE estimator can be expressed as
\begin{align}\label{MMSE_Estimator_SIMO_Basic}
\mathbbmss{E}\left\{\left.x\right|\textbf{y}\right\}=\int xp\left(x|\textbf{y},\mathsf{snr},{\textbf{h}}\right){\rm{d}}x
=\frac{\sum_{i=1}^{M}p_ix_i
\frac{1}{\pi^N}\exp\left(-{\left\|\textbf{y}-\sqrt{\mathsf{snr}}x_i{\textbf{h}}\right\|^2}\right)}
{\sum_{i=1}^{M}p_i
\frac{1}{\pi^N}\exp\left(-{\left\|\textbf{y}-\sqrt{\mathsf{snr}}x_i{\textbf{h}}\right\|^2}\right)}\triangleq{\mathcal{S}}_{\text{simo}}\left(\textbf{y}\right).
\end{align}
Accordingly, the MMSE can be calculated as
\begin{align}
{\mathsf{mmse}}_{\text{simo}}\left(\mathsf{snr};{\textbf{h}}\right)=\left\|{\textbf{h}}\right\|^2\mathbbmss{E}\left\{\left.\left|x-{\mathcal{S}}_{\text{simo}}\left(\textbf{y}\right)\right|^2\right|\mathsf{snr},{\textbf{h}}\right\},
\end{align}
where the expectation is taken over $\left(x,\textbf{y}\right)$. Therefore, the MMSE can be calculated as
\begin{align}
{\mathsf{mmse}}_{\text{simo}}\left(\mathsf{snr};{\textbf{h}}\right)&=
\left\|{\textbf{h}}\right\|^2\int\int\left|x-{\mathcal{S}}_{\text{simo}}\left(\textbf{y}\right)\right|^2p\left(\left.x,\textbf{y}\right|\mathsf{snr},{\textbf{h}}\right)
{\rm{d}}x{\rm{d}}\textbf{y}\\
&=\left\|{\textbf{h}}\right\|^2
\int\sum_{i=1}^{M}p_i\left|x_i-{\mathcal{S}}_{\text{simo}}\left(\textbf{y}\right)\right|^2
\frac{1}{\pi^N}\exp\left(-{\left\|\textbf{y}-\sqrt{\mathsf{snr}}x_i{\textbf{h}}\right\|^2}\right){\rm{d}}\textbf{y}.
\end{align}
Following similar steps in obtaining \eqref{MMSE_SISO_Complex}, we can get
\begin{align}
{\mathsf{mmse}}_{\text{simo}}\left(\mathsf{snr};{\textbf{h}}\right)=\left\|{\textbf{h}}\right\|^2
\left(
1-\frac{1}{\pi^N}\int\frac{\left|\sum_{i=1}^{M}p_ix_i
\exp\left(-{\left\|\textbf{y}-\sqrt{\mathsf{snr}}x_i{\textbf{h}}\right\|^2}\right)\right|^2}
{\sum_{i=1}^{M}p_i
\exp\left(-{\left\|\textbf{y}-\sqrt{\mathsf{snr}}x_i{\textbf{h}}\right\|^2}\right)}{\rm{d}}\textbf{y}
\right).
\end{align}
In fact, the MMSE estimator shown in \eqref{MMSE_Estimator_SIMO_Basic} can be written as
\begin{align}
{\mathcal{S}}_{\text{simo}}\left(\textbf{y}\right)&=
\frac{\sum_{i=1}^{M}p_ix_i
\exp\left(
-{\textbf{y}}^{\dag}{\textbf{y}}-\mathsf{snr}x_i^{\dag}x_i{\textbf{h}}^{\dag}{\textbf{h}}+2\sqrt{\mathsf{snr}}\Re\left\{x_i{\textbf{y}}^{\dag}{\textbf{h}}\right\}
\right)}
{\sum_{i=1}^{M}p_i
\exp\left(
-{\textbf{y}}^{\dag}{\textbf{y}}-\mathsf{snr}x_i^{\dag}x_i{\textbf{h}}^{\dag}{\textbf{h}}+2\sqrt{\mathsf{snr}}\Re\left\{x_i{\textbf{y}}^{\dag}{\textbf{h}}\right\}
\right)}\\
&=\frac{\sum_{i=1}^{M}p_ix_i
\exp\left(
-\mathsf{snr}x_i^{\dag}x_i\left\|{\textbf{h}}\right\|^2+2\sqrt{\mathsf{snr}}\Re\left\{x_i{\textbf{y}}^{\dag}{\textbf{h}}\right\}
\right)}
{\sum_{i=1}^{M}p_i
\exp\left(
-\mathsf{snr}x_i^{\dag}x_i\left\|{\textbf{h}}\right\|^2+2\sqrt{\mathsf{snr}}\Re\left\{x_i{\textbf{y}}^{\dag}{\textbf{h}}\right\}
\right)}\\
&=\frac{\sum_{i=1}^{M}p_ix_i
\exp\left(-\frac{{\textbf{y}}^{\dag}{\textbf{h}}}{\left\|{\textbf{h}}\right\|}\frac{{\textbf{h}}^{\dag}{\textbf{y}}}{\left\|{\textbf{h}}\right\|}
-\mathsf{snr}x_i^{\dag}x_i\left\|{\textbf{h}}\right\|^2+2\sqrt{\mathsf{snr}}\left\|{\textbf{h}}\right\|\Re\left\{x_i\frac{{\textbf{y}}^{\dag}{\textbf{h}}}{\left\|{\textbf{h}}\right\|}\right\}
\right)}
{\sum_{i=1}^{M}p_i
\exp\left(-\frac{{\textbf{y}}^{\dag}{\textbf{h}}}{\left\|{\textbf{h}}\right\|}\frac{{\textbf{h}}^{\dag}{\textbf{y}}}{\left\|{\textbf{h}}\right\|}
-\mathsf{snr}x_i^{\dag}x_i\left\|{\textbf{h}}\right\|^2+2\sqrt{\mathsf{snr}}\left\|{\textbf{h}}\right\|\Re\left\{x_i\frac{{\textbf{y}}^{\dag}{\textbf{h}}}{\left\|{\textbf{h}}\right\|}\right\}
\right)}\\
&=\frac{\sum_{i=1}^{M}p_ix_i
\frac{1}{\pi}\exp\left(-{\left|\frac{{\textbf{y}}^{\dag}{\textbf{h}}}{\left\|{\textbf{h}}\right\|}-\sqrt{\mathsf{snr}}\left\|{\textbf{h}}\right\|x_i\right|^2}\right)}
{\sum_{i=1}^{M}p_i
\frac{1}{\pi}\exp\left(-{\left|\frac{{\textbf{y}}^{\dag}{\textbf{h}}}{\left\|{\textbf{h}}\right\|}-\sqrt{\mathsf{snr}}\left\|{\textbf{h}}\right\|x_i\right|^2}\right)}.
\label{MMSE_Estimator_SIMO_Trans}
\end{align}
It is worth noting that $\textbf{h}$ is fixed and thus $\frac{{\textbf{y}}^{\dag}{\textbf{h}}}{\left\|{\textbf{h}}\right\|}$ is fixed once $\textbf{y}$ is given. Hence, the MMSE estimator in \eqref{MMSE_Estimator_SIMO_Trans} can be treated as the MMSE estimator of the following complex SISO channel:
\begin{align}
\tilde{y}=\frac{{\textbf{y}}^{\dag}{\textbf{h}}}{\left\|{\textbf{h}}\right\|}=\sqrt{\mathsf{snr}}\left\|\textbf{h}\right\|x+\frac{{\textbf{z}}^{\dag}{\textbf{h}}}{\left\|{\textbf{h}}\right\|},
\end{align}
where $\tilde{z}=\frac{{\textbf{z}}^{\dag}{\textbf{h}}}{\left\|{\textbf{h}}\right\|}\sim{\mathcal{CN}}\left(0,1\right)$. Therefore, the corresponding MMSE can be expressed as
\begin{align}
\int\int\left|x-\frac{\sum_{i=1}^{M}p_ix_i
\frac{1}{\pi}\exp\left(-{\left|\tilde{y}-\sqrt{\mathsf{snr}}\left\|{\textbf{h}}\right\|x_i\right|^2}\right)}
{\sum_{i=1}^{M}p_i
\frac{1}{\pi}\exp\left(-{\left|\tilde{y}-\sqrt{\mathsf{snr}}\left\|{\textbf{h}}\right\|x_i\right|^2}\right)}\right|^2p\left(\left.x,\tilde{y}\right|\mathsf{snr},{\textbf{h}}\right){\rm{d}}x{\rm{d}}\tilde{y}.
\end{align}
We comment that the result of the above integral equals
\begin{align}
&\int\int\left|x-{\mathcal{S}}_{\text{simo}}\left(\textbf{y}\right)\right|^2p\left(\left.x,\textbf{y}\right|\mathsf{snr},{\textbf{h}}\right)
{\rm{d}}x{\rm{d}}\textbf{y}\\
&=\int\int\left|x-\frac{\sum_{i=1}^{M}p_ix_i
\frac{1}{\pi}\exp\left(-{\left|\frac{{\textbf{y}}^{\dag}{\textbf{h}}}{\left\|{\textbf{h}}\right\|}-\sqrt{\mathsf{snr}}\left\|{\textbf{h}}\right\|x_i\right|^2}\right)}
{\sum_{i=1}^{M}p_i
\frac{1}{\pi}\exp\left(-{\left|\frac{{\textbf{y}}^{\dag}{\textbf{h}}}{\left\|{\textbf{h}}\right\|}-\sqrt{\mathsf{snr}}\left\|{\textbf{h}}\right\|x_i\right|^2}\right)}
\right|^2p\left(\left.x,\textbf{y}\right|\mathsf{snr},{\textbf{h}}\right)
{\rm{d}}x{\rm{d}}\textbf{y}.
\end{align}
The reason for this lies in that the random variable $\textbf{y}$ influences the integrand function via the term $\frac{{\textbf{y}}^{\dag}{\textbf{h}}}{\left\|{\textbf{h}}\right\|}$. Leveraging the fact of
\begin{equation}
\begin{split}
&\int\int\left|x-{\mathcal{S}}_{\text{simo}}\left(\textbf{y}\right)\right|^2p\left(\left.x,\textbf{y}\right|\mathsf{snr},{\textbf{h}}\right)
{\rm{d}}x{\rm{d}}\textbf{y}\\
&=\int\int\left|x-\frac{\sum_{i=1}^{M}p_ix_i
\frac{1}{\pi}\exp\left(-{\left|\tilde{y}-\sqrt{\mathsf{snr}}\left\|{\textbf{h}}\right\|x_i\right|^2}\right)}
{\sum_{i=1}^{M}p_i
\frac{1}{\pi}\exp\left(-{\left|\tilde{y}-\sqrt{\mathsf{snr}}\left\|{\textbf{h}}\right\|x_i\right|^2}\right)}\right|^2p\left(\left.x,\tilde{y}\right|\mathsf{snr},{\textbf{h}}\right){\rm{d}}x{\rm{d}}\tilde{y},
\end{split}
\end{equation}
we can arrive at the following result
\begin{align}
&{\mathsf{mmse}}_{\text{simo}}\left(\mathsf{snr};{\textbf{h}}\right)=
\left\|{\textbf{h}}\right\|^2\int\int\left|x-{\mathcal{S}}_{\text{simo}}\left(\textbf{y}\right)\right|^2p\left(\left.x,\textbf{y}\right|\mathsf{snr},{\textbf{h}}\right)
{\rm{d}}x{\rm{d}}\textbf{y}\\
&=\left\|{\textbf{h}}\right\|^2\int\int\left|x-\frac{\sum_{i=1}^{M}p_ix_i
\frac{1}{\pi}\exp\left(-{\left|\tilde{y}-\sqrt{\mathsf{snr}}\left\|{\textbf{h}}\right\|x_i\right|^2}\right)}
{\sum_{i=1}^{M}p_i
\frac{1}{\pi}\exp\left(-{\left|\tilde{y}-\sqrt{\mathsf{snr}}\left\|{\textbf{h}}\right\|x_i\right|^2}\right)}\right|^2p\left(\left.x,\tilde{y}\right|\mathsf{snr},{\textbf{h}}\right){\rm{d}}x{\rm{d}}\tilde{y}\\
&=\left\|{\textbf{h}}\right\|^2{\mathsf{mmse}}_{\text{siso}}\left(\mathsf{snr}\left\|{\textbf{h}}\right\|^2\right)\\
&=\left\|{\textbf{h}}\right\|^2\left(1-\frac{1}{\pi}\int\frac{\left|\sum_{i=1}^{M}p_ix_i
\exp\left(-{\left|y-\sqrt{\mathsf{snr}}\left\|{\textbf{h}}\right\|x_i\right|^2}\right)\right|^2}
{\sum_{i=1}^{M}p_i
\exp\left(-{\left|y-\sqrt{\mathsf{snr}}\left\|{\textbf{h}}\right\|x_i\right|^2}\right)}{\rm{d}}y\right).
\end{align}
In the following, we specialize the above expression to the case of BPSK. In this case, we have
\begin{align}
{\mathsf{mmse}}_{\text{simo}}\left(\mathsf{snr};{\textbf{h}}\right)=\left\|{\textbf{h}}\right\|^2\left(1-\int_{-\infty}^{+\infty}
\tanh\left(2\sqrt{\mathsf{snr}}\left\|{\textbf{h}}\right\|a\right)
\frac{\exp\left(-{\left(a-\sqrt{\mathsf{snr}}\left\|{\textbf{h}}\right\|\right)^2}\right)}{\sqrt{\pi}}
{\rm{d}}a\right).
\end{align}

\section{Multiple-Input Multiple-Output (MIMO)}
Finally, let us consider a MIMO channel given by
\begin{equation}
{\textbf{y}}=\sqrt{\mathsf{snr}}{\textbf{H}}{\textbf{x}}+{\textbf{n}},
\end{equation}
where ${\textbf{H}}\in{\mathbb{C}}^{N\times N_{\text{t}}}$ denotes the channel matrix, ${\textbf{n}}\sim{\mathcal{CN}}\left({\textbf{0}},{\textbf{I}}_{N}\right)$ denotes the complex additive white Gaussian noise, $\textbf{x}\in{\mathbb{C}}^{N_{\text{t}}\times 1}$ denotes the transmitted signal vector satisfying ${\mathbb{E}}\left\{{\textbf{x}}{\textbf{x}}^{\dag}\right\}=\frac{1}{N_{\text{t}}}{\textbf{I}}_{N_{\text{t}}}$, and $\mathsf{snr}$ denotes the SNR. Furthermore, we assume that $\textbf{x}$ is taken from an infinite alphabet $\mathcal{X}=\left\{{\textbf{x}}_1,\cdots,{\textbf{x}}_{M}\right\}$ containing $M$ elements with probability $\Pr\left({\textbf{x}}={\textbf{x}}_l\right)=p_l$. The MSE achieved by estimator, $f\left(\textbf{y}\right)\in{\mathbb{C}}^{N_{\text{t}}\times1}$, is given by
\begin{align}
{\mathsf{mse}}&=\mathbbmss{E}\left\{\left.\left\|{\textbf{H}}\left(\textbf{x}-f\left(\textbf{y}\right)\right)\right\|^2\right|\textbf{y}\right\}\\
&=\mathbbmss{E}\left\{\left.{\textbf{x}}^{\dag}{\textbf{H}}^{\dag}{\textbf{H}}{\textbf{x}}\right|\textbf{y}\right\}+
f^{\dag}\left(\textbf{y}\right){\textbf{H}}^{\dag}{\textbf{H}}f\left(\textbf{y}\right)-f^{\dag}\left(\textbf{y}\right){\textbf{H}}^{\dag}{\textbf{H}}
\mathbbmss{E}\left\{\left.\textbf{x}\right|\textbf{y}\right\}-
\mathbbmss{E}\left\{\left.\textbf{x}^{\dag}\right|\textbf{y}\right\}{\textbf{H}}^{\dag}{\textbf{H}}f\left(\textbf{y}\right)
.
\end{align}
Upon calculating the complex gradient of ${\mathsf{mse}}$ with respect to $f\left(\textbf{y}\right)$ and then setting it to zero, we can obtain the MMSE estimator as follows
\begin{align}
f^{\star}\left(\textbf{y}\right)=\arg\min_{f\left(\textbf{y}\right)}=\mathbbmss{E}\left\{\left.\textbf{x}\right|\textbf{y}\right\}.
\end{align}
Consequently, the corresponding MMSE can be expressed as
\begin{align}
{\mathsf{mmse}}_{\text{mimo}}\left(\mathsf{snr};{\textbf{h}}\right)
&={\mathbbmss{E}}\left\{\left.\left\|
{\textbf{H}}{\textbf{x}}-{\textbf{H}}{\mathbbmss{E}}\left\{\left.{\textbf{x}}\right|{\textbf{y}}\right\}\right\|^2
\right|{\textbf{H}}\right\}\\
&=
\int\int\left\|
{\textbf{H}}{\textbf{x}}-{\textbf{H}}{\mathbbmss{E}}\left\{\left.{\textbf{x}}\right|{\textbf{y}}\right\}\right\|^2p\left(\left.\textbf{x},\textbf{y}\right|\mathsf{snr},{\textbf{H}}\right){\rm{d}}\textbf{x}{\rm{d}}\textbf{y}.
\end{align}
In this complex-value MIMO channel, the MI-MMSE relationship satisfies
\begin{align}
\frac{{\rm{d}}}{{\rm{d}}{\mathsf{snr}}}{\mathsf{I}}_{\text{mimo}}\left(\mathsf{snr};{\textbf{H}}\right)=
{\mathsf{mmse}}_{\text{mimo}}\left(\mathsf{snr};{\textbf{H}}\right),
\end{align}
where
\begin{align}
{\mathsf{I}}_{\text{mimo}}\left(\mathsf{snr};{\textbf{H}}\right)={\mathbbmss{E}}\left\{\left.\log\left(\frac{p\left({\textbf{x}},{\textbf{y}}|{\mathsf{snr}},{\textbf{H}}\right)}
{p\left({\textbf{x}}|{\mathsf{snr}},{\textbf{H}}\right)p\left({\textbf{y}}|{\mathsf{snr}},{\textbf{H}}\right)}\right)\right|{\mathsf{snr}},{\textbf{H}}\right\},
\end{align}
denotes the mutual information of this channel.
\begin{equation}
\begin{split}
p\left(\left.\textbf{x},\textbf{y}\right|\mathsf{snr},{\textbf{H}}\right)=\sum_{i=1}^{M}p_i\delta\left(\textbf{x}-\textbf{x}_i\right)
\frac{1}{\pi^N}\exp\left(-{\left\|\textbf{y}-\sqrt{\mathsf{snr}}{\textbf{H}}\textbf{x}_i\right\|^2}\right),
\end{split}
\end{equation}
and the PDF of $\textbf{y}$ for a given $\mathsf{snr}$ can be given as
\begin{align}
p\left(\left.\textbf{y}\right|\mathsf{snr},{\textbf{H}}\right)=\sum_{i=1}^{M}p_i
\frac{1}{\pi^N}\exp\left(-{\left\|\textbf{y}-\sqrt{\mathsf{snr}}{\textbf{H}}\textbf{x}_i\right\|^2}\right).
\end{align}
As a result, the MMSE estimator can be expressed as
\begin{align}\label{MMSE_Estimator_SIMO_Basic}
\mathbbmss{E}\left\{\left.\textbf{x}\right|\textbf{y}\right\}=\int \textbf{x}p\left(\textbf{x}|\textbf{y},\mathsf{snr},{\textbf{H}}\right){\rm{d}}x
=\frac{\sum_{i=1}^{M}p_i\textbf{x}_i
\frac{1}{\pi^N}\exp\left(-{\left\|\textbf{y}-\sqrt{\mathsf{snr}}{\textbf{H}}\textbf{x}_i\right\|^2}\right)}
{\sum_{i=1}^{M}p_i
\frac{1}{\pi^N}\exp\left(-{\left\|\textbf{y}-\sqrt{\mathsf{snr}}{\textbf{H}}\textbf{x}_i\right\|^2}\right)}\triangleq{\mathcal{S}}_{\text{mimo}}\left(\textbf{y}\right).
\end{align}
Accordingly, the MMSE can be calculated as
\begin{align}
{\mathsf{mmse}}_{\text{mimo}}\left(\mathsf{snr};{\textbf{H}}\right)={\mathbbmss{E}}\left\{\left.\left\|
{\textbf{H}}{\textbf{x}}-{\textbf{H}}{\mathcal{S}}_{\text{mimo}}\left(\textbf{y}\right)\right\|^2
\right|{\textbf{H}}\right\},
\end{align}
where the expectation is taken over $\left(\textbf{x},\textbf{y}\right)$. Therefore, the MMSE can be calculated as
\begin{align}
{\mathsf{mmse}}_{\text{simo}}\left(\mathsf{snr};{\textbf{h}}\right)&=
\int\int\left\|
{\textbf{H}}{\textbf{x}}-{\textbf{H}}{\mathcal{S}}_{\text{mimo}}\left(\textbf{y}\right)\right\|^2p\left(\left.\textbf{x},\textbf{y}\right|\mathsf{snr},{\textbf{H}}\right)
{\rm{d}}\textbf{x}{\rm{d}}\textbf{y}\\
&=
\int\sum_{i=1}^{M}p_i\left\|
{\textbf{H}}{\textbf{x}}_i-{\textbf{H}}{\mathcal{S}}_{\text{mimo}}\left(\textbf{y}\right)\right\|^2
\frac{1}{\pi^N}\exp\left(-{\left\|\textbf{y}-\sqrt{\mathsf{snr}}{\textbf{H}}\textbf{x}_i\right\|^2}\right){\rm{d}}\textbf{y},
\end{align}
which can be further written as
\begin{equation}
\begin{split}
&{\mathsf{mmse}}_{\text{mimo}}\left(\mathsf{snr};{\textbf{H}}\right)\\
&={\mathsf{tr}}\left({\textbf{H}}\left[\sum_{i=1}^{M}p_i
\int\left({\textbf{x}}_i-{\mathcal{S}}_{\text{mimo}}\left(\textbf{y}\right)\right)
\left({\textbf{x}}_i-{\mathcal{S}}_{\text{mimo}}\left(\textbf{y}\right)\right)^{\dag}
\frac{\exp\left(-{\left\|\textbf{y}-\sqrt{\mathsf{snr}}{\textbf{H}}\textbf{x}_i\right\|^2}\right)}{\pi^N}{\rm{d}}\textbf{y}\right]{\textbf{H}}^{\dag}\right).
\end{split}
\end{equation}
Following similar steps in obtaining \eqref{MMSE_SISO_Complex}, we can get
\begin{equation}\label{MMSE_MIMO_Final}
\begin{split}
&{\mathsf{mmse}}_{\text{mimo}}\left(\mathsf{snr};{\textbf{H}}\right)\\
&={\mathsf{tr}}\left({\textbf{H}}\left[{\textbf{I}}_N-
\int\frac{\left(\sum\limits_{i=1}^{M}p_i\textbf{x}_i
\frac{\exp\left(-{\left\|\textbf{y}-\sqrt{\mathsf{snr}}{\textbf{H}}\textbf{x}_i\right\|^2}\right)}{\pi^N}\right)
\left(\sum\limits_{i=1}^{M}p_i\textbf{x}_i
\frac{\exp\left(-{\left\|\textbf{y}-\sqrt{\mathsf{snr}}{\textbf{H}}\textbf{x}_i\right\|^2}\right)}{\pi^N}\right)^{\dag}
}
{\sum_{i=1}^{M}p_i
\frac{1}{\pi^N}\exp\left(-{\left\|\textbf{y}-\sqrt{\mathsf{snr}}{\textbf{H}}\textbf{x}_i\right\|^2}\right)}
{\rm{d}}\textbf{y}\right]{\textbf{H}}^{\dag}\right).
\end{split}
\end{equation}

We note that the MMSE of MIMO channels presents a complicated form, which makes the associated analyses even harder. To that end, we present two bounds for the MMSE shown in \eqref{MMSE_MIMO_Final}. We first consider the upper bound. Since the MSE achieved by the MMSE estimator is the minimum, the MSE achieved by other estimator is larger than that achieved by the MMSE estimator and thus serves as an upper bound of the MMSE. Particularly, we consider the maximum-likelihood (ML) estimator which is expressed as
\begin{align}
{\mathcal{S}}_{\text{ML}}\left({\textbf{y}}\right)=\arg\min_{{\textbf{x}}\in{\mathcal{X}}}{\left\|{\textbf{y}}-\sqrt{\mathsf{snr}}{\textbf{H}}\textbf{x}\right\|^2},
\end{align}
thereby
\begin{align}
&{\mathsf{mmse}}_{\text{mimo}}\left(\mathsf{snr};{\textbf{H}}\right)={\mathbbmss{E}}\left\{\left.\left\|
{\textbf{H}}{\textbf{x}}-{\textbf{H}}{\mathcal{S}}_{\text{mimo}}\left(\textbf{y}\right)\right\|^2
\right|{\textbf{H}}\right\}\leq {\mathbbmss{E}}\left\{\left.\left\|
{\textbf{H}}{\textbf{x}}-{\textbf{H}}{\mathcal{S}}_{\text{ML}}\left(\textbf{y}\right)\right\|^2
\right|{\textbf{H}}\right\}\\
&=\frac{1}{M}\sum\nolimits_{i=1}^{M}\sum\nolimits_{j=1,j\neq i}^{M}{\mathbbmss{E}}\left\{\left.\left\|{\textbf{H}}\left({\textbf{x}}_i-{\textbf{x}}_j\right)\right\|^2\right|{\textbf{x}}={\textbf{x}}_i,{\textbf{y}}\in{\mathcal{V}}_j\right\}\Pr\left(\left.{\textbf{y}}\in{\mathcal{V}}_j\right|{\textbf{x}}={\textbf{x}}_i\right)\\
&=\frac{1}{M}\sum\nolimits_{i=1}^{M}\sum\nolimits_{j=1,j\neq i}^{M}\left\|{\textbf{H}}\left({\textbf{x}}_i-{\textbf{x}}_j\right)\right\|^2\Pr\left(\left.{\textbf{y}}\in{\mathcal{V}}_j\right|{\textbf{x}}={\textbf{x}}_i\right),
\end{align}
where ${\mathcal{V}}_j$ is the Voronoi region for ${\textbf{H}}{\textbf{x}}_j$ in the received constellation. Note that we have assumed that all the constellation vectors are equally likely. Consequently, we can obtain
\begin{align}
{\mathsf{mmse}}_{\text{mimo}}\left(\mathsf{snr};{\textbf{H}}\right)&\leq 
\frac{1}{M}\sum_{i=1}^{M}\sum_{j=1,j\neq i}^{M}\left\|{\textbf{H}}\left({\textbf{x}}_i-{\textbf{x}}_j\right)\right\|^2\Pr\left(\left.{\textbf{y}}\in{\mathcal{V}}_j\right|{\textbf{x}}={\textbf{x}}_i\right)\\
&\leq\frac{1}{M}\sum_{i=1}^{M}\sum_{j=1,j\neq i}^{M}\left\|{\textbf{H}}\left({\textbf{x}}_i-{\textbf{x}}_j\right)\right\|^2\Pr\left(\left.{\textbf{y}}\notin{\mathcal{V}}_{ij}\right|{\textbf{x}}={\textbf{x}}_i\right)\\
&=\frac{1}{M}\sum_{i=1}^{M}\sum_{j=1,j\neq i}^{M}\left\|{\textbf{H}}\left({\textbf{x}}_i-{\textbf{x}}_j\right)\right\|^2Q\left(\sqrt{\frac{{\mathsf{snr}}}{2}\left\|{\textbf{H}}\left({\textbf{x}}_i-{\textbf{x}}_j\right)\right\|^2}\right),
\end{align}
where ${\mathcal{V}}_{ij}$ is the Voronoi region for ${\textbf{H}}{\textbf{x}}_i$, when we assume that only ${\textbf{x}}_i$ and ${\textbf{x}}_j$ have been transmitted. It is worth noting that the probability when ${\textbf{y}}$ does not fall into the Voronoi region for ${\textbf{H}}{\textbf{x}}_i$ when ${\textbf{x}}_i$ is sent equals the probability of the project of the noise vector alone the direction of ${\textbf{H}}{\textbf{x}}_j-{\textbf{H}}{\textbf{x}}_i$ being larger than half the distance between $\sqrt{\mathsf{snr}}{\textbf{H}}{\textbf{x}}_j$ and $\sqrt{\mathsf{snr}}{\textbf{H}}{\textbf{x}}_i$, namely
\begin{align}
\Pr\left(\left.{\textbf{y}}\notin{\mathcal{V}}_{ij}\right|{\textbf{x}}={\textbf{x}}_i\right)=
\Pr\left({\mathsf{p}}>\frac{1}{2}\sqrt{{\mathsf{snr}}}\left\|{\textbf{H}}\left({\textbf{x}}_i-{\textbf{x}}_j\right)\right\|\right),
\end{align}
where ${\textbf{n}}={\textbf{y}}-\sqrt{\mathsf{snr}}{\textbf{H}}{\textbf{x}}_i\sim{\mathcal{CN}}\left({\textbf{0}},{\textbf{I}}_N\right)$, ${\textbf{z}}=\left[\Re\left\{{\textbf{n}}\right\}^{\mathsf{T}},\Im\left\{{\textbf{n}}\right\}^{\mathsf{T}}\right]^{\mathsf{T}}\sim{\mathcal{CN}}\left({\textbf{0}},\frac{1}{2}{\textbf{I}}_{2N}\right)$,
${\mathsf{p}}={\textbf{z}}^{\mathsf{T}}\frac{{\textbf{x}}_{ij}}{\left\|{\textbf{x}}_{ij}\right\|}$ denotes the project of $\textbf{z}$ in the direction of $\textbf{x}_{ij}$, and $\textbf{x}_{ij}=\left[\Re\left\{{\textbf{H}}{\textbf{x}}_j-{\textbf{H}}{\textbf{x}}_i\right\}^{\mathsf{T}},\Im\left\{{\textbf{H}}{\textbf{x}}_j-{\textbf{H}}{\textbf{x}}_i\right\}^{\mathsf{T}}\right]^{\mathsf{T}}$. Therefore, ${\mathsf{p}}\sim{\mathcal{CN}}\left(0,\frac{1}{2}\right)$, and the cumulative distribution function (CDF) of $\mathsf{p}$ is given by
\begin{align}
F_{\mathsf{p}}\left(x\right)=\int_{-\infty}^{x}\frac{1}{\sqrt{2\pi\times \frac{1}{2}}}{\text{e}}^{-\frac{x^2}{2\times \frac{1}{2}}}{\rm{d}}x=
\int_{-\infty}^{x}\frac{1}{\sqrt{\pi}}{\text{e}}^{-{x^2}}{\rm{d}}x.
\end{align} 
It follows that 
\begin{align}
\Pr\left({\mathsf{p}}>\frac{1}{2}\sqrt{{\mathsf{snr}}}\left\|{\textbf{H}}\left({\textbf{x}}_i-{\textbf{x}}_j\right)\right\|\right)&=
\int_{\frac{\sqrt{{\mathsf{snr}}}\left\|{\textbf{H}}\left({\textbf{x}}_i-{\textbf{x}}_j\right)\right\|}{2}}^{\infty}\frac{1}{\sqrt{\pi}}{\text{e}}^{-{x^2}}{\rm{d}}x\\
&=
\int_{\frac{\sqrt{{\mathsf{snr}}}\left\|{\textbf{H}}\left({\textbf{x}}_i-{\textbf{x}}_j\right)\right\|}{\sqrt{2}}}^{\infty}\frac{1}{\sqrt{2\pi}}{\text{e}}^{-\frac{x^2}{2}}{\rm{d}}x\\
&=Q\left(\sqrt{\frac{{\mathsf{snr}}}{2}\left\|{\textbf{H}}\left({\textbf{x}}_i-{\textbf{x}}_j\right)\right\|^2}\right),
\end{align}
where $Q\left(x\right)=\frac{1}{\sqrt{2\pi}}\int_{x}^{\infty}{\text{e}}^{-\frac{x^2}{2}}{\rm{d}}x$ is the Q-function. Consequently, we can obtain
\begin{align}
{\mathsf{mmse}}_{\text{mimo}}\left(\mathsf{snr};{\textbf{H}}\right)\leq \frac{1}{M}\sum_{i=1}^{M}\sum_{j=1,j\neq i}^{M}\left\|{\textbf{H}}\left({\textbf{x}}_i-{\textbf{x}}_j\right)\right\|^2Q\left(\sqrt{\frac{{\mathsf{snr}}}{2}\left\|{\textbf{H}}\left({\textbf{x}}_i-{\textbf{x}}_j\right)\right\|^2}\right).
\end{align}

Having characterized the upper bound of the MMSE, we now turn our attention to the lower bound of the MMSE. To derive the lower bound, we can postulate the existence of a genie that informs the estimator pair of constellation vectors: if the transmitted constellation vector is $\textbf{x}_i$, then with probability $\frac{1}{M-1}$ the genie gives the pair $\left\{{\textbf{x}}_i,{\textbf{x}}_j\right\}$. More specifically, this genie will tell the estimator that it should only consider the points $\left\{{\textbf{x}}_i,{\textbf{x}}_j\right\}$ equiprobably. Once the estimator knows that it only needs to consider the points $\left\{{\textbf{x}}_i,{\textbf{x}}_j\right\}$ equiprobably, the corresponding MMSE estimator is thus given by
\begin{align}\label{MMSE_Estimator_Genie_MIMO_Basic}
\frac{\frac{1}{2}\textbf{x}_i
\frac{1}{\pi^N}\exp\left(-{\left\|\textbf{y}-\sqrt{\mathsf{snr}}{\textbf{H}}\textbf{x}_i\right\|^2}\right)+
\frac{1}{2}\textbf{x}_j
\frac{1}{\pi^N}\exp\left(-{\left\|\textbf{y}-\sqrt{\mathsf{snr}}{\textbf{H}}\textbf{x}_j\right\|^2}\right)}
{\frac{1}{2}
\frac{1}{\pi^N}\exp\left(-{\left\|\textbf{y}-\sqrt{\mathsf{snr}}{\textbf{H}}\textbf{x}_i\right\|^2}\right)+
\frac{1}{2}
\frac{1}{\pi^N}\exp\left(-{\left\|\textbf{y}-\sqrt{\mathsf{snr}}{\textbf{H}}\textbf{x}_j\right\|^2}\right)}
\triangleq{\mathcal{S}}_{\text{genie}}\left(\textbf{y},\left\{{\textbf{x}}_i,{\textbf{x}}_j\right\}\right).
\end{align}
As a result, the MSE achieved by this genie-aided estimator is given by
\begin{align}
\frac{1}{M\left(M-1\right)}\sum_{i=1}^{M}\sum_{j=1,j\neq i}^{M}{\mathbbmss{E}}\left\{\left.\left\|
{\textbf{H}}\left({\textbf{x}}_i-{\mathcal{S}}_{\text{genie}}\left(\textbf{y},\left\{{\textbf{x}}_i,{\textbf{x}}_j\right\}\right)\right)
\right\|^2\right|{\textbf{x}}={\textbf{x}}_i\right\}.
\end{align}
Since the genie-aided estimator knows more prior information than the conventional MMSE estimator as well as using the MMSE estimator to deal with this prior information, its achieved MSE is lower than the MMSE. Thus, we conclude that the MMSE is lower bounded by
\begin{align}
{\mathsf{mmse}}_{\text{mimo}}\left(\mathsf{snr};{\textbf{H}}\right)\geq\frac{1}{M\left(M-1\right)}\sum_{i=1}^{M}\sum_{j=1,j\neq i}^{M}{\mathbbmss{E}}\left\{\left.\left\|
{\textbf{H}}\left({\textbf{x}}_i-{\mathcal{S}}_{\text{genie}}\left(\textbf{y},\left\{{\textbf{x}}_i,{\textbf{x}}_j\right\}\right)\right)
\right\|^2\right|{\textbf{x}}={\textbf{x}}_i\right\}.
\end{align}
In the sequel, we try to characterize ${\mathbbmss{E}}\left\{\left.\left\|
{\textbf{H}}\left({\textbf{x}}_i-{\mathcal{S}}_{\text{genie}}\left(\textbf{y},\left\{{\textbf{x}}_i,{\textbf{x}}_j\right\}\right)\right)
\right\|^2\right|{\textbf{x}}={\textbf{x}}_i\right\}$. Particularly, it is worth noting that
\begin{align}
{\textbf{y}}=\sqrt{\mathsf{snr}}{\textbf{H}}{\textbf{x}}+{\textbf{n}},~{\textbf{x}}\in\left\{{\textbf{x}}_i,{\textbf{x}}_j\right\},
\end{align}
is equivalent to
\begin{align}
{\textbf{y}}=\sqrt{\mathsf{snr}}\frac{{\textbf{H}}\left({\textbf{x}}_i-{\textbf{x}}_j\right)}{2}{{x}}+
\sqrt{\mathsf{snr}}\frac{{\textbf{H}}\left({\textbf{x}}_i+{\textbf{x}}_j\right)}{2}
+{\textbf{n}},~{{x}}\in\left\{+1,-1\right\}.
\end{align} 
Based on this fact, we find that the following relationship also holds
\begin{align}
&{\mathcal{S}}_{\text{genie}}\left(\textbf{y},\left\{{\textbf{x}}_i,{\textbf{x}}_j\right\}\right)=
\frac{\textbf{x}_i
\exp\left(-{\left\|\textbf{y}-\sqrt{\mathsf{snr}}{\textbf{H}}\textbf{x}_i\right\|^2}\right)+
\textbf{x}_j
\exp\left(-{\left\|\textbf{y}-\sqrt{\mathsf{snr}}{\textbf{H}}\textbf{x}_j\right\|^2}\right)}
{\exp\left(-{\left\|\textbf{y}-\sqrt{\mathsf{snr}}{\textbf{H}}\textbf{x}_i\right\|^2}\right)+
\exp\left(-{\left\|\textbf{y}-\sqrt{\mathsf{snr}}{\textbf{H}}\textbf{x}_j\right\|^2}\right)}\\
&=\frac{
\exp\left(-{\left\|\textbf{y}-\sqrt{\mathsf{snr}}\frac{{\textbf{H}}\left({\textbf{x}}_i+{\textbf{x}}_j\right)}{2}
-\sqrt{\mathsf{snr}}\frac{{\textbf{H}}\left({\textbf{x}}_i-{\textbf{x}}_j\right)}{2}\right\|^2}\right)-
\exp\left(-{\left\|\textbf{y}-\sqrt{\mathsf{snr}}\frac{{\textbf{H}}\left({\textbf{x}}_i+{\textbf{x}}_j\right)}{2}
+\sqrt{\mathsf{snr}}\frac{{\textbf{H}}\left({\textbf{x}}_i-{\textbf{x}}_j\right)}{2}\right\|^2}\right)}
{\exp\left(-{\left\|\textbf{y}-\sqrt{\mathsf{snr}}\frac{{\textbf{H}}\left({\textbf{x}}_i+{\textbf{x}}_j\right)}{2}
-\sqrt{\mathsf{snr}}\frac{{\textbf{H}}\left({\textbf{x}}_i-{\textbf{x}}_j\right)}{2}\right\|^2}\right)+
\exp\left(-{\left\|\textbf{y}-\sqrt{\mathsf{snr}}\frac{{\textbf{H}}\left({\textbf{x}}_i+{\textbf{x}}_j\right)}{2}
+\sqrt{\mathsf{snr}}\frac{{\textbf{H}}\left({\textbf{x}}_i-{\textbf{x}}_j\right)}{2}\right\|^2}\right)}\nonumber\\
&\times\frac{\left({\textbf{x}}_i-{\textbf{x}}_j\right)}{2}+\frac{\left({\textbf{x}}_i+{\textbf{x}}_j\right)}{2}=
\mathcal{S}\left(\textbf{y},\left\{{\textbf{x}}_i,{\textbf{x}}_j\right\}\right)\frac{\left({\textbf{x}}_i-{\textbf{x}}_j\right)}{2}+\frac{\left({\textbf{x}}_i+{\textbf{x}}_j\right)}{2}.
\end{align}
It is worth noting that $\mathcal{S}\left(\textbf{y},\left\{{\textbf{x}}_i,{\textbf{x}}_j\right\}\right)$ can be treated as the MMSE estimator of 
\begin{align}
\tilde{\textbf{y}}={\textbf{y}}-
\sqrt{\mathsf{snr}}\frac{{\textbf{H}}\left({\textbf{x}}_i+{\textbf{x}}_j\right)}{2}=
\sqrt{\mathsf{snr}}\frac{{\textbf{H}}\left({\textbf{x}}_i-{\textbf{x}}_j\right)}{2}{{x}}
+{\textbf{n}},~{{x}}\in\left\{+1,-1\right\}.
\end{align}
Consequently, 
\begin{align}
&{\mathbbmss{E}}\left\{\left.\left\|
{\textbf{H}}\left({\textbf{x}}_i-{\mathcal{S}}_{\text{genie}}\left(\textbf{y},\left\{{\textbf{x}}_i,{\textbf{x}}_j\right\}\right)\right)
\right\|^2\right|{\textbf{x}}={\textbf{x}}_i\right\}\nonumber\\
&={\mathbbmss{E}}_{\textbf{y}}\left\{\left\|
{\textbf{H}}\left({\textbf{x}}_i-\frac{\textbf{x}_i
\exp\left(-{\left\|\textbf{y}-\sqrt{\mathsf{snr}}{\textbf{H}}\textbf{x}_i\right\|^2}\right)+
\textbf{x}_j
\exp\left(-{\left\|\textbf{y}-\sqrt{\mathsf{snr}}{\textbf{H}}\textbf{x}_j\right\|^2}\right)}
{\exp\left(-{\left\|\textbf{y}-\sqrt{\mathsf{snr}}{\textbf{H}}\textbf{x}_i\right\|^2}\right)+
\exp\left(-{\left\|\textbf{y}-\sqrt{\mathsf{snr}}{\textbf{H}}\textbf{x}_j\right\|^2}\right)}\right)
\right\|^2\right\}\\
&={\mathbbmss{E}}_{\textbf{y}}\left\{\left\|
{\textbf{H}}\left({\textbf{x}}_i-\mathcal{S}\left(\textbf{y},\left\{{\textbf{x}}_i,{\textbf{x}}_j\right\}\right)\frac{\left({\textbf{x}}_i-{\textbf{x}}_j\right)}{2}
-\frac{\left({\textbf{x}}_i+{\textbf{x}}_j\right)}{2}\right)
\right\|^2\right\}\\
&={\mathbbmss{E}}_{\textbf{y}}\left\{\left\|\frac{{\textbf{H}}\left({\textbf{x}}_i-{\textbf{x}}_j\right)}{2}
\left(1-\mathcal{S}\left(\textbf{y},\left\{{\textbf{x}}_i,{\textbf{x}}_j\right\}\right)\right)
\right\|^2\right\},
\end{align}
and it follows that
\begin{align}
{\mathbbmss{E}}\left\{\left.\left\|
{\textbf{H}}\left({\textbf{x}}_i-{\mathcal{S}}_{\text{genie}}\left(\textbf{y},\left\{{\textbf{x}}_i,{\textbf{x}}_j\right\}\right)\right)
\right\|^2\right|{\textbf{x}}={\textbf{x}}_i\right\}=
\left\|\frac{{\textbf{H}}\left({\textbf{x}}_i-{\textbf{x}}_j\right)}{2}\right\|^2
{\mathbbmss{E}}_{\textbf{y}}\left\{\left|
1-\mathcal{S}\left(\textbf{y},\left\{{\textbf{x}}_i,{\textbf{x}}_j\right\}\right)
\right|^2\right\}.
\end{align}
Since the random variable $\textbf{y}$ influences the statistics of $\mathcal{S}\left(\textbf{y},\left\{{\textbf{x}}_i,{\textbf{x}}_j\right\}\right)$ via the term $\textbf{y}-\sqrt{\mathsf{snr}}\frac{{\textbf{H}}\left({\textbf{x}}_i+{\textbf{x}}_j\right)}{2}$, we can get
\begin{align}
{\mathbbmss{E}}_{\textbf{y}}\left\{\left|
1-\mathcal{S}\left(\textbf{y},\left\{{\textbf{x}}_i,{\textbf{x}}_j\right\}\right)
\right|^2\right\}=
{\mathbbmss{E}}_{\tilde{\textbf{y}}}\left\{\left|
1-\mathcal{S}\left(\textbf{y},\left\{{\textbf{x}}_i,{\textbf{x}}_j\right\}\right)
\right|^2\right\}.
\end{align}
As mentioned, $\mathcal{S}\left(\textbf{y},\left\{{\textbf{x}}_i,{\textbf{x}}_j\right\}\right)$ can be treated as the MMSE estimator of an SIMO channel with BPSK inputs. Using our previous derived results, we find that $\mathcal{S}\left(\textbf{y},\left\{{\textbf{x}}_i,{\textbf{x}}_j\right\}\right)$ can be aslo treated as the MMSE estimator of an SISO channel with channel gain $\sqrt{\mathsf{snr}}\left\|\frac{{\textbf{H}}\left({\textbf{x}}_i-{\textbf{x}}_j\right)}{2}\right\|$ and BPSK inputs. Therefore, 
\begin{align}
&{\mathbbmss{E}}_{\tilde{\textbf{y}}}\left\{\left|
1-\mathcal{S}\left(\textbf{y},\left\{{\textbf{x}}_i,{\textbf{x}}_j\right\}\right)
\right|^2\right\}\nonumber\\
&={\mathbbmss{E}}_{\tilde{{y}}}\left\{\left|
1-\frac{\exp\left(-{\left|\tilde{y}-\sqrt{\mathsf{snr}}\left\|\frac{{\textbf{H}}\left({\textbf{x}}_i-{\textbf{x}}_j\right)}{2}\right\|\right|^2}\right)-
\exp\left(-{\left|\tilde{y}+\sqrt{\mathsf{snr}}\left\|\frac{{\textbf{H}}\left({\textbf{x}}_i-{\textbf{x}}_j\right)}{2}\right\|\right|^2}\right)
}
{\exp\left(-{\left|\tilde{y}-\sqrt{\mathsf{snr}}\left\|\frac{{\textbf{H}}\left({\textbf{x}}_i-{\textbf{x}}_j\right)}{2}\right\|\right|^2}\right)+
\exp\left(-{\left|\tilde{y}+\sqrt{\mathsf{snr}}\left\|\frac{{\textbf{H}}\left({\textbf{x}}_i-{\textbf{x}}_j\right)}{2}\right\|\right|^2}\right)}
\right|^2\right\},
\end{align}
where $\tilde{{y}}=\frac{{\tilde{\textbf{y}}}^{\dag}{\frac{{\textbf{H}}\left({\textbf{x}}_i-{\textbf{x}}_j\right)}{2}}}
{\left\|\frac{{\textbf{H}}\left({\textbf{x}}_i-{\textbf{x}}_j\right)}{2}\right\|}=\sqrt{\mathsf{snr}}\left\|\frac{{\textbf{H}}\left({\textbf{x}}_i-{\textbf{x}}_j\right)}{2}\right\|x+\tilde{n}$ with $\tilde{n}\sim{\mathcal{CN}}\left(0,1\right)$. For simplicity, denote $C_g=\sqrt{\mathsf{snr}}\left\|\frac{{\textbf{H}}\left({\textbf{x}}_i-{\textbf{x}}_j\right)}{2}\right\|$. Consequently, we can obtain
\begin{align}
&{\mathbbmss{E}}_{\tilde{\textbf{y}}}\left\{\left|
1-\mathcal{S}\left(\textbf{y},\left\{{\textbf{x}}_i,{\textbf{x}}_j\right\}\right)
\right|^2\right\}={\mathbbmss{E}}_{\tilde{{y}}}\left\{\left|
1-\frac{\exp\left(-{\left|\tilde{y}-C_g\right|^2}\right)-
\exp\left(-{\left|\tilde{y}+C_g\right|^2}\right)
}
{\exp\left(-{\left|\tilde{y}-C_g\right|^2}\right)+
\exp\left(-{\left|\tilde{y}+C_g\right|^2}\right)}
\right|^2\right\},\\
&=\frac{1}{2\pi}\int_{\mathbb{C}}{\frac{4
\exp\left(-2{\left|\tilde{y}+C_g\right|^2}\right)
}
{\exp\left(-{\left|\tilde{y}-C_g\right|^2}\right)+
\exp\left(-{\left|\tilde{y}+C_g\right|^2}\right)}}{\rm{d}}\tilde{y}.
\end{align}
Changing the variable $\tilde{y}\rightarrow-\tilde{y}$, we can obtain
\begin{align}
&\frac{1}{2\pi}\int_{\mathbb{C}}{\frac{4
\exp\left(-2{\left|\tilde{y}+C_g\right|^2}\right)
}
{\exp\left(-{\left|\tilde{y}-C_g\right|^2}\right)+
\exp\left(-{\left|\tilde{y}+C_g\right|^2}\right)}}{\rm{d}}\tilde{y}\nonumber\\&=
\frac{1}{2\pi}\int_{\mathbb{C}}{\frac{4
\exp\left(-2{\left|-\tilde{y}+C_g\right|^2}\right)
}
{\exp\left(-{\left|-\tilde{y}-C_g\right|^2}\right)+
\exp\left(-{\left|-\tilde{y}+C_g\right|^2}\right)}}{\rm{d}}\tilde{y}\\
&=\frac{1}{2\pi}\int_{\mathbb{C}}{\frac{4
\exp\left(-2{\left|\tilde{y}-C_g\right|^2}\right)
}
{\exp\left(-{\left|\tilde{y}-C_g\right|^2}\right)+
\exp\left(-{\left|\tilde{y}+C_g\right|^2}\right)}}{\rm{d}}\tilde{y}.
\end{align}
Furthermore, we find that
\begin{align}
&{\mathbbmss{E}}_{\tilde{\textbf{y}}}\left\{\left|
-1-\mathcal{S}\left(\textbf{y},\left\{{\textbf{x}}_i,{\textbf{x}}_j\right\}\right)
\right|^2\right\}={\mathbbmss{E}}_{\tilde{{y}}}\left\{\left|
-1-\frac{\exp\left(-{\left|\tilde{y}-C_g\right|^2}\right)-
\exp\left(-{\left|\tilde{y}+C_g\right|^2}\right)
}
{\exp\left(-{\left|\tilde{y}-C_g\right|^2}\right)+
\exp\left(-{\left|\tilde{y}+C_g\right|^2}\right)}
\right|^2\right\},\\
&=\frac{1}{2\pi}\int_{\mathbb{C}}{\frac{4
\exp\left(-2{\left|\tilde{y}-C_g\right|^2}\right)
}
{\exp\left(-{\left|\tilde{y}-C_g\right|^2}\right)+
\exp\left(-{\left|\tilde{y}+C_g\right|^2}\right)}}{\rm{d}}\tilde{y}.
\end{align}
Taken together, we have
\begin{align}
{\mathbbmss{E}}_{\tilde{\textbf{y}}}\left\{\left|
1-\mathcal{S}\left(\textbf{y},\left\{{\textbf{x}}_i,{\textbf{x}}_j\right\}\right)
\right|^2\right\}={\mathbbmss{E}}_{\tilde{\textbf{y}}}\left\{\left|
-1-\mathcal{S}\left(\textbf{y},\left\{{\textbf{x}}_i,{\textbf{x}}_j\right\}\right)
\right|^2\right\}.
\end{align}
Actually, 
\begin{align}
&\frac{1}{2}{\mathbbmss{E}}_{\tilde{\textbf{y}}}\left\{\left|
1-\mathcal{S}\left(\textbf{y},\left\{{\textbf{x}}_i,{\textbf{x}}_j\right\}\right)
\right|^2\right\}+\frac{1}{2}{\mathbbmss{E}}_{\tilde{\textbf{y}}}\left\{\left|
-1-\mathcal{S}\left(\textbf{y},\left\{{\textbf{x}}_i,{\textbf{x}}_j\right\}\right)
\right|^2\right\}\nonumber\\
&={\mathbbmss{E}}_{x\in\left\{-1,+1\right\},\tilde{\textbf{y}}}\left\{\left|
x-\mathcal{S}\left(\textbf{y},\left\{{\textbf{x}}_i,{\textbf{x}}_j\right\}\right)
\right|^2\right\},
\end{align}
which is the MMSE achieved by BPSK signals. Using previous results, we have
\begin{align}
&{\mathbbmss{E}}_{x\in\left\{-1,+1\right\},\tilde{\textbf{y}}}\left\{\left|
x-\mathcal{S}\left(\textbf{y},\left\{{\textbf{x}}_i,{\textbf{x}}_j\right\}\right)
\right|^2\right\}\nonumber\\
&=1-\int_{-\infty}^{+\infty}
\tanh\left(2C_ga\right)
\frac{\exp\left(-{\left(a-C_g\right)^2}\right)}{\sqrt{\pi}}
{\rm{d}}a\\
&={\mathbbmss{E}}_{\tilde{\textbf{y}}}\left\{\left|
1-\mathcal{S}\left(\textbf{y},\left\{{\textbf{x}}_i,{\textbf{x}}_j\right\}\right)
\right|^2\right\}={\mathbbmss{E}}_{\tilde{\textbf{y}}}\left\{\left|
-1-\mathcal{S}\left(\textbf{y},\left\{{\textbf{x}}_i,{\textbf{x}}_j\right\}\right)
\right|^2\right\}.
\end{align}
In summary, we can obtain
\begin{align}
&{\mathbbmss{E}}\left\{\left.\left\|
{\textbf{H}}\left({\textbf{x}}_i-{\mathcal{S}}_{\text{genie}}\left(\textbf{y},\left\{{\textbf{x}}_i,{\textbf{x}}_j\right\}\right)\right)
\right\|^2\right|{\textbf{x}}={\textbf{x}}_i\right\}=\left\|\frac{{\textbf{H}}\left({\textbf{x}}_i-{\textbf{x}}_j\right)}{2}\right\|^2\nonumber\\
&\times\left[1-\int_{-\infty}^{+\infty}
\tanh\left(2\sqrt{\mathsf{snr}}\left\|\frac{{\textbf{H}}\left({\textbf{x}}_i-{\textbf{x}}_j\right)}{2}\right\|a\right)
\frac{\exp\left(-{\left(a-\sqrt{\mathsf{snr}}\left\|\frac{{\textbf{H}}\left({\textbf{x}}_i-{\textbf{x}}_j\right)}{2}\right\|\right)^2}\right)}{\sqrt{\pi}}
{\rm{d}}a\right],
\end{align}
and it follows that
\begin{align}
&{\mathsf{mmse}}_{\text{mimo}}\left(\mathsf{snr};{\textbf{H}}\right)\geq\frac{1}{M\left(M-1\right)}\sum_{i=1}^{M}\sum_{j=1,j\neq i}^{M}{\mathbbmss{E}}\left\{\left.\left\|
{\textbf{H}}\left({\textbf{x}}_i-{\mathcal{S}}_{\text{genie}}\left(\textbf{y},\left\{{\textbf{x}}_i,{\textbf{x}}_j\right\}\right)\right)
\right\|^2\right|{\textbf{x}}={\textbf{x}}_i\right\}\\
&=\frac{1}{M\left(M-1\right)}\sum_{i=1}^{M}\sum_{j=1,j\neq i}^{M}\left\|\frac{{\textbf{H}}\left({\textbf{x}}_i-{\textbf{x}}_j\right)}{2}\right\|^2\nonumber\\
&\times\left[1-\int_{-\infty}^{+\infty}
\tanh\left(2\sqrt{\mathsf{snr}}\left\|\frac{{\textbf{H}}\left({\textbf{x}}_i-{\textbf{x}}_j\right)}{2}\right\|a\right)
\frac{\exp\left(-{\left(a-\sqrt{\mathsf{snr}}\left\|\frac{{\textbf{H}}\left({\textbf{x}}_i-{\textbf{x}}_j\right)}{2}\right\|\right)^2}\right)}{\sqrt{\pi}}
{\rm{d}}a\right].
\end{align}
Upon defining $d_{i,j}=\left\|{\textbf{H}}{\textbf{x}}_i-{\textbf{H}}{\textbf{x}}_j\right\|$, we arrive at the following conclusion:
\begin{align}
{\mathsf{mmse}}_{\text{mimo},l}\left(\mathsf{snr};{\textbf{H}}\right)\leq
{\mathsf{mmse}}_{\text{mimo}}\left(\mathsf{snr};{\textbf{H}}\right)\leq
{\mathsf{mmse}}_{\text{mimo},u}\left(\mathsf{snr};{\textbf{H}}\right),
\end{align}
where
\begin{align}
&{\mathsf{mmse}}_{\text{mimo},l}\left(\mathsf{snr};{\textbf{H}}\right)=\frac{1}{M\left(M-1\right)}\nonumber\\
&\times\sum_{i=1}^{M}\sum_{j=1,j\neq i}^{M}\frac{d_{i,j}^2}{4}
\left[1-\int_{-\infty}^{+\infty}
\tanh\left(\sqrt{\mathsf{snr}}d_{i,j}a\right)
\frac{\exp\left(-{\left(a-\frac{\sqrt{\mathsf{snr}}}{2}d_{i,j}\right)^2}\right)}{\sqrt{\pi}}
{\rm{d}}a\right]
\end{align}
and
\begin{align}
{\mathsf{mmse}}_{\text{mimo},u}\left(\mathsf{snr};{\textbf{H}}\right)=\frac{1}{M}\sum_{i=1}^{M}\sum_{j=1,j\neq i}^{M}d_{i,j}^2
Q\left(\sqrt{\frac{{\mathsf{snr}}}{2}d_{i,j}^2}\right).
\end{align}

Now, we consider the mutual information of the MIMO system, which is defined as
\begin{align}
{\mathsf{I}}_{\text{mimo}}\left(\mathsf{snr};{\textbf{H}}\right)&={\mathbbmss{E}}_{{\textbf{x}},{\textbf{y}}}\left\{\left.\log\left(\frac{p\left({\textbf{x}},{\textbf{y}}|{\mathsf{snr}},{\textbf{H}}\right)}
{p\left({\textbf{x}}|{\mathsf{snr}},{\textbf{H}}\right)p\left({\textbf{y}}|{\mathsf{snr}},{\textbf{H}}\right)}\right)\right|{\mathsf{snr}},{\textbf{H}}\right\}\\
&={\mathbbmss{E}}\left\{\left.\log\left(\frac{p\left({\textbf{y}}|{\textbf{x}},{\mathsf{snr}},{\textbf{H}}\right)}
{p\left({\textbf{y}}|{\mathsf{snr}},{\textbf{H}}\right)}\right)\right|{\mathsf{snr}},{\textbf{H}}\right\}.
\end{align}
Particularly, we have
\begin{align}
&p\left({\textbf{y}}|{\textbf{x}},{\mathsf{snr}},{\textbf{H}}\right)=
\frac{1}{\pi^N}\exp\left(-{\left\|\textbf{y}-\sqrt{\mathsf{snr}}{\textbf{H}}\textbf{x}\right\|^2}\right),\\
&p\left(\left.\textbf{y}\right|\mathsf{snr},{\textbf{H}}\right)=\sum_{i=1}^{M}p_i
\frac{1}{\pi^N}\exp\left(-{\left\|\textbf{y}-\sqrt{\mathsf{snr}}{\textbf{H}}\textbf{x}_i\right\|^2}\right).
\end{align}
It follows that
\begin{align}
{\mathsf{I}}_{\text{mimo}}\left(\mathsf{snr};{\textbf{H}}\right)&=
{\mathbbmss{E}}_{{\textbf{x}},{\textbf{y}}}\left\{\left.\log\left(\frac{\exp\left(-{\left\|\textbf{y}-\sqrt{\mathsf{snr}}{\textbf{H}}\textbf{x}\right\|^2}\right)}
{\sum_{i=1}^{M}p_i\exp\left(-{\left\|\textbf{y}-\sqrt{\mathsf{snr}}{\textbf{H}}\textbf{x}_i\right\|^2}\right)}\right)\right|{\mathsf{snr}},{\textbf{H}}\right\}\\
&={\mathbbmss{E}}_{{\textbf{x}},{\textbf{y}}}\left\{\left.\log\left({\exp\left(-{\left\|\textbf{y}-\sqrt{\mathsf{snr}}{\textbf{H}}\textbf{x}\right\|^2}\right)}
\right)\right|{\mathsf{snr}},{\textbf{H}}\right\}\nonumber\\
&-{\mathbbmss{E}}_{{\textbf{x}},{\textbf{y}}}\left\{\left.\log\left(
{\sum_{i=1}^{M}p_i\exp\left(-{\left\|\textbf{y}-\sqrt{\mathsf{snr}}{\textbf{H}}\textbf{x}_i\right\|^2}\right)}\right)\right|{\mathsf{snr}},{\textbf{H}}\right\}\\
&=-{\mathbbmss{E}}_{{\textbf{x}},{\textbf{y}}}\left\{\left.{{\left\|\textbf{y}-\sqrt{\mathsf{snr}}{\textbf{H}}\textbf{x}\right\|^2}}
\right|{\mathsf{snr}},{\textbf{H}}\right\}\nonumber\\
&-{\mathbbmss{E}}_{{\textbf{y}}}\left\{\left.\log\left(
{\sum_{i=1}^{M}p_i\exp\left(-{\left\|\textbf{y}-\sqrt{\mathsf{snr}}{\textbf{H}}\textbf{x}_i\right\|^2}\right)}\right)\right|{\mathsf{snr}},{\textbf{H}}\right\}.
\end{align}
Furthermore, 
\begin{align}
&{\mathbbmss{E}}_{{\textbf{x}},{\textbf{y}}}\left\{\left.{{\left\|\textbf{y}-\sqrt{\mathsf{snr}}{\textbf{H}}\textbf{x}\right\|^2}}
\right|{\mathsf{snr}},{\textbf{H}}\right\}\nonumber\\
&=\sum_{i=1}^{M}p_i\int_{{\mathbb{C}}^N}\left\|\textbf{y}-\sqrt{\mathsf{snr}}{\textbf{H}}\textbf{x}_i\right\|^2
\frac{1}{\pi^N}\exp\left(-{\left\|\textbf{y}-\sqrt{\mathsf{snr}}{\textbf{H}}\textbf{x}_i\right\|^2}\right){\rm{d}}{\textbf{y}}\\
&=\sum_{i=1}^{M}p_i{\mathbbmss{E}}\left\{\left\|{\textbf{n}}\right\|^2\right\}=\sum_{i=1}^{M}p_iN=N.
\end{align}
We than consider the value of ${\mathbbmss{E}}_{{\textbf{y}}}\left\{\left.\log\left(
{\sum_{i=1}^{M}p_i\exp\left(-{\left\|\textbf{y}-\sqrt{\mathsf{snr}}{\textbf{H}}\textbf{x}_i\right\|^2}\right)}\right)\right|{\mathsf{snr}},{\textbf{H}}\right\}$, which is characterized as
\begin{align}
&{\mathbbmss{E}}_{{\textbf{y}}}\left\{\left.\log\left(
{\sum_{i=1}^{M}p_i\exp\left(-{\left\|\textbf{y}-\sqrt{\mathsf{snr}}{\textbf{H}}\textbf{x}_i\right\|^2}\right)}\right)\right|{\mathsf{snr}},{\textbf{H}}\right\}\nonumber\\
&=\sum_{i=1}^{M}\int_{{\mathbb{C}}^N}p_i
\frac{1}{\pi^N}\exp\left(-{\left\|\textbf{y}-\sqrt{\mathsf{snr}}{\textbf{H}}\textbf{x}_i\right\|^2}\right)\log\left(
{\sum_{j=1}^{M}p_i\exp\left(-{\left\|\textbf{y}-\sqrt{\mathsf{snr}}{\textbf{H}}\textbf{x}_j\right\|^2}\right)}\right){\rm{d}}{\textbf{y}}\\
&=\sum_{i=1}^{M}\int_{{\mathbb{C}}^N}p_i
\frac{1}{\pi^N}\exp\left(-{\left\|\textbf{y}\right\|^2}\right)\log\left(
{\sum_{j=1}^{M}p_i\exp\left(-{\left\|\textbf{y}+\sqrt{\mathsf{snr}}{\textbf{H}}\textbf{x}_i-\sqrt{\mathsf{snr}}{\textbf{H}}\textbf{x}_j\right\|^2}\right)}\right){\rm{d}}{\textbf{y}}\\
&=\sum_{i=1}^{M}p_i\int_{{\mathbb{C}}^N}
\frac{1}{\pi^N}\exp\left(-{\left\|\textbf{n}\right\|^2}\right)\log\left(
{\sum_{j=1}^{M}p_i\exp\left(-{\left\|\textbf{n}+\sqrt{\mathsf{snr}}{\textbf{H}}\textbf{x}_i-\sqrt{\mathsf{snr}}{\textbf{H}}\textbf{x}_j\right\|^2}\right)}\right){\rm{d}}{\textbf{n}}\\
&=\sum_{i=1}^{M}p_i{\mathbbmss{E}}_{{\textbf{n}}}\left\{\left.\log\left(
{\sum_{i=1}^{M}p_i\exp\left(-{\left\|\textbf{n}+\sqrt{\mathsf{snr}}{\textbf{H}}\textbf{x}_i-\sqrt{\mathsf{snr}}{\textbf{H}}\textbf{x}_j\right\|^2}\right)}\right)\right|{\mathsf{snr}},{\textbf{H}}\right\}.
\end{align}
Taken together, the mutual information can be written as
\begin{align}
{\mathsf{I}}_{\text{mimo}}\left(\mathsf{snr};{\textbf{H}}\right)=-N-
\sum_{i=1}^{M}p_i{\mathbbmss{E}}_{{\textbf{n}}}\left\{\left.\log\left(
{\sum_{i=1}^{M}p_i\exp\left(-{\left\|\textbf{n}+\sqrt{\mathsf{snr}}{\textbf{H}}\left(\textbf{x}_i-\textbf{x}_j\right)\right\|^2}\right)}\right)\right|{\mathsf{snr}},{\textbf{H}}\right\}.
\end{align} 
We then consider a special case when $p_i=\frac{1}{M}$. In this case, we can obtain 
\begin{align}
{\mathsf{I}}_{\text{mimo}}\left(\mathsf{snr};{\textbf{H}}\right)=\log{M}-N-
\sum_{i=1}^{M}\frac{1}{M}{\mathbbmss{E}}_{{\textbf{n}}}\left\{\left.\log\left(
{\sum_{i=1}^{M}\exp\left(-{\left\|\textbf{n}+\sqrt{\mathsf{snr}}{\textbf{H}}\left(\textbf{x}_i-\textbf{x}_j\right)\right\|^2}\right)}\right)\right|{\mathsf{snr}},{\textbf{H}}\right\}.
\end{align} 
In the following, let us consider two special cases of $\mathsf{snr}\rightarrow0$ and $\mathsf{snr}\rightarrow\infty$. When $\mathsf{snr}\rightarrow0$, we have
\begin{align}
{\mathsf{I}}_{\text{mimo}}\left(0;{\textbf{H}}\right)&=\log{M}-N-
\sum_{i=1}^{M}\frac{1}{M}{\mathbbmss{E}}_{{\textbf{n}}}\left\{\left.\log\left(
{\sum_{i=1}^{M}\exp\left(-{\left\|\textbf{n}\right\|^2}\right)}\right)\right|{\mathsf{snr}},{\textbf{H}}\right\}\\
&=\log{M}-N-\sum_{i=1}^{M}\frac{1}{M}\left(\log{M}-N\right)=0.
\end{align}
When $\mathsf{snr}\rightarrow\infty$, there exists no noise and the channel model degrades into
\begin{align}
{\textbf{y}}=\sqrt{\mathsf{snr}}{\textbf{H}}{\textbf{x}}.
\end{align}
In this case, the mutual information can be written as
\begin{align}
{\mathsf{I}}_{\text{mimo}}\left(\infty;{\textbf{H}}\right)&=H\left({\textbf{x}}\right)+H\left({\textbf{y}}\right)-H\left({\textbf{x}};{\textbf{y}}\right)\\
&=\sum_{i=1}^{M}p_i\log{\frac{1}{p_i}}+H\left({\textbf{y}}\right)-H\left({\textbf{x}};{\textbf{y}}\right),
\end{align} 
where $H\left(\cdot\right)$ is the entropy function. If ${{\textbf{x}}_i\neq {\textbf{x}}_j}\Longleftrightarrow{{\textbf{y}}_i\neq {\textbf{y}}_j}$ (actually, this condition holds for probability 1 for fading channels), then we have
\begin{align}
H\left({\textbf{x}}\right)=H\left({\textbf{y}}\right)=H\left({\textbf{x}};{\textbf{y}}\right),
\end{align}
which yields
\begin{align}
{\mathsf{I}}_{\text{mimo}}\left(\infty;{\textbf{H}}\right)=H\left({\textbf{x}}\right)=\sum_{i=1}^{M}p_i\log{\frac{1}{p_i}}.
\end{align}
Based on the famous MI-MMSE relationship, we have
\begin{align}
\frac{{\rm{d}}}{{\rm{d}}{\mathsf{snr}}}{\mathsf{I}}_{\text{mimo}}\left(\mathsf{snr};{\textbf{H}}\right)=
{\mathsf{mmse}}_{\text{mimo}}\left(\mathsf{snr};{\textbf{H}}\right),
\end{align}
which suggest that
\begin{align}
{\mathsf{I}}_{\text{mimo}}\left(\mathsf{snr};{\textbf{H}}\right)=
\int_{0}^{x}{\mathsf{mmse}}_{\text{mimo}}\left(x;{\textbf{H}}\right){\rm{d}}x={\mathsf{I}}_{\text{mimo}}\left(\infty;{\textbf{H}}\right)
-\int_{x}^{\infty}{\mathsf{mmse}}_{\text{mimo}}\left(x;{\textbf{H}}\right){\rm{d}}x.
\end{align}
Leveraging the derived bounds of the MMSE, we can obtain the bounds of mutual information as follows
\begin{align}
{\mathsf{I}}_{\text{mimo}}\left(\infty;{\textbf{H}}\right)
-\int_{x}^{\infty}{\mathsf{mmse}}_{\text{mimo},u}\left(x;{\textbf{H}}\right){\rm{d}}x
&\leq{\mathsf{I}}_{\text{mimo}}\left(\mathsf{snr};{\textbf{H}}\right)\nonumber\\
&\leq
{\mathsf{I}}_{\text{mimo}}\left(\infty;{\textbf{H}}\right)
-\int_{x}^{\infty}{\mathsf{mmse}}_{\text{mimo},l}\left(x;{\textbf{H}}\right){\rm{d}}x.
\end{align}

We now continue to evaluate the asymptotic behaviour of the average mutual information in terms of channel fading. Particularly, the average mutual information of the MIMO system can be written as 
\begin{align}
\bar{\mathsf{I}}_{\text{mimo}}\left(\mathsf{snr}\right)={\mathbbmss{E}}_{\textbf{H}}\left\{{\mathsf{I}}_{\text{mimo}}\left(\mathsf{snr};{\textbf{H}}\right)\right\}.
\end{align}
As investigated earlier, the MMSE function has the following properties
\begin{align}
{\mathsf{mmse}}_{\text{mimo},l}\left(\mathsf{snr};{\textbf{H}}\right)\leq
{\mathsf{mmse}}_{\text{mimo}}\left(\mathsf{snr};{\textbf{H}}\right)\leq
{\mathsf{mmse}}_{\text{mimo},u}\left(\mathsf{snr};{\textbf{H}}\right),
\end{align}
where
\begin{align}
{\mathsf{mmse}}_{\text{mimo},l}\left(\mathsf{snr};{\textbf{H}}\right)
=\frac{1}{M}\frac{1}{M-1}\sum_{i=1}^{M}\sum_{j=1,j\neq i}^{M}d_{i,j}^2f_l\left(\mathsf{snr} d_{i,j}^2\right)
\end{align}
with
\begin{align}
f_l\left(x\right)=\frac{1}{4}
\left[1-\int_{-\infty}^{+\infty}
\tanh\left(\sqrt{x}a\right)
\frac{\exp\left(-{\left(a-\frac{\sqrt{x}}{2}\right)^2}\right)}{\sqrt{\pi}}
{\rm{d}}a\right],
\end{align}
and where
\begin{align}
{\mathsf{mmse}}_{\text{mimo},u}\left(\mathsf{snr};{\textbf{H}}\right)=\frac{1}{M}\sum_{i=1}^{M}\sum_{j=1,j\neq i}^{M}d_{i,j}^2f_u\left(\mathsf{snr} d_{i,j}^2\right)
\end{align}
with
\begin{align}
f_u\left(x\right)=Q\left(\sqrt{\frac{{x}}{2}}\right)
\end{align}

Thus, we have
\begin{align}
&\bar{\mathsf{I}}_{\text{mimo}}\left(\mathsf{snr}\right)
\geq {\mathbbmss{E}}_{\textbf{H}}\left\{{\mathsf{I}}_{\text{mimo}}\left(\infty;{\textbf{H}}\right)\right\}
-{\mathbbmss{E}}_{\textbf{H}}\left\{\int_{x}^{\infty}{\mathsf{mmse}}_{\text{mimo},u}\left(x;{\textbf{H}}\right){\rm{d}}x\right\}\\
&=
{\mathbbmss{E}}_{\textbf{H}}\left\{{\mathsf{I}}_{\text{mimo}}\left(\infty;{\textbf{H}}\right)\right\}-\frac{1}{M}\frac{1}{M-1}\sum_{i=1}^{M}\sum_{j=1,j\neq i}^{M}\int_{0}^{\infty} \int_{\mathsf{snr}}^{\infty}f_{i,j}\left(y\right)
yf_u\left(xy\right){\rm{d}}x{\rm{d}}y\\
&=
{\mathbbmss{E}}_{\textbf{H}}\left\{{\mathsf{I}}_{\text{mimo}}\left(\infty;{\textbf{H}}\right)\right\}-\frac{1}{M}\frac{1}{M-1}\sum_{i=1}^{M}\sum_{j=1,j\neq i}^{M}\int_{0}^{\infty} f_{i,j}\left(\frac{y}{\mathsf{snr}}\right)\frac{y}{\mathsf{snr}}\int_{y}^{\infty}
f_u\left(x\right){\rm{d}}x{\rm{d}}y,
\end{align}
where $f_{i,j}\left(\cdot\right)$ is the PDF of $d_{i,j}^2=\left\|{\textbf{H}}{\textbf{x}}_i-{\textbf{H}}{\textbf{x}}_j\right\|^2$. As $\mathsf{snr}\rightarrow\infty$, we have $f_{i,j}\left(\frac{y}{\mathsf{snr}}\right)={\mathcal{O}}\left(\frac{1}{{\mathsf{snr}}^{N-1}}\right)$.
Hence, we have
\begin{equation}
\bar{\mathsf{I}}_{\text{mimo}}\left(\mathsf{snr}\right)\geq{\mathbbmss{E}}_{\textbf{H}}\left\{{\mathsf{I}}_{\text{mimo}}\left(\infty;{\textbf{H}}\right)\right\}
-{\mathcal{O}}\left(\frac{1}{{\mathsf{snr}}^{N}}\right).
\end{equation}
Following similar steps, we can also get
\begin{equation}
\bar{\mathsf{I}}_{\text{mimo}}\left(\mathsf{snr}\right)\leq{\mathbbmss{E}}_{\textbf{H}}\left\{{\mathsf{I}}_{\text{mimo}}\left(\infty;{\textbf{H}}\right)\right\}
-{\mathcal{O}}\left(\frac{1}{{\mathsf{snr}}^{N}}\right).
\end{equation}
Taken together, we can obtain
\begin{align}
\bar{\mathsf{I}}_{\text{mimo}}\left(\mathsf{snr}\right)={\mathbbmss{E}}_{\textbf{H}}\left\{{\mathsf{I}}_{\text{mimo}}\left(\infty;{\textbf{H}}\right)\right\}
+{\mathcal{O}}\left(\frac{1}{{\mathsf{snr}}^{N}}\right),
\end{align}
which suggests that the average mutual information statures to its maximum ${\mathbbmss{E}}_{\textbf{H}}\left\{{\mathsf{I}}_{\text{mimo}}\left(\infty;{\textbf{H}}\right)\right\}$ as $\mathsf{snr}$ increases and the diversity order is given by $N$.

\section{Conclusion}
Two closed-form bounds were derived for the MMSE and MI of a complex-valued MIMO channel with arbitrary input signals. Based on these two bound, we discuss the asymptotic average mutual information of MIMO fading channels in the high-SNR region. Theoretical analyses indicate that the diversity order of the MIMO system equals the receive antenna number.

\end{document}